\documentclass[titlepage]{article}%
\usepackage{amsmath}
\usepackage{cite}%
\usepackage{amsfonts}%
\usepackage{amssymb}%
\usepackage{graphicx}
\newtheorem{theorem}{Theorem}

\newtheorem{example}{Example}

\newenvironment{proof}[1][Proof]{\noindent\textbf{#1.} }{\ \rule{0.5em}{0.5em}}
\begin{document}

\title{Network-based analysis of stochastic SIR epidemic models with random and
proportionate mixing}
\author{Eben Kenah*, James M. Robins\\Departments of Epidemiology and Biostatistics\\Harvard School of Public Health\\677 Huntington Ave., Boston MA 02115\\*Correspondence to: ekenah@hsph.harvard.edu}
\date{April-September 2006\\
Revised January-September, 2007}
\maketitle

\begin{abstract}
In this paper, we outline the theory of epidemic percolation networks and
their use in the analysis of stochastic SIR epidemic models on undirected
contact networks. \ We then show how the same theory can be used to analyze
stochastic SIR models with random and proportionate mixing. \ The epidemic
percolation networks for these models are purely directed because undirected
edges disappear in the limit of a large population. \ In a series of
simulations, we show that epidemic percolation networks accurately predict the
mean outbreak size and probability and final size of an epidemic for a variety
of epidemic models in homogeneous and heterogeneous populations. \ Finally, we
show that epidemic percolation networks can be used to re-derive classical
results from several different areas of infectious disease epidemiology. \ In
an appendix, we show that an epidemic percolation network can be defined for
any time-homogeneous stochastic SIR model in a closed population and prove
that the distribution of outbreak sizes given the infection of any given node
in the SIR\ model is identical to the distribution of its out-component sizes
in the corresponding probability space of epidemic percolation networks. \ We
conclude that the theory of percolation on semi-directed networks provides a
very general framework for the analysis of stochastic SIR models in closed populations.

\end{abstract}

\section{Introduction}

In an important paper, M. E. J. Newman studied a network-based
Susceptible-Infectious-Removed (SIR) epidemic model in which infection is
transmitted through a network of contacts between individuals \cite{Newman1}.
\ The contact network itself was a random undirected network with an arbitrary
degree distribution of the form studied by Newman, Strogatz, and Watts
\cite{NSW}. \ Given the degree distribution, these networks are maximally
random. \ Thus, they have no small loops in the limit of a large population
\cite{NSW,Boguna,Meyers}. \ 

In the SIR model from \cite{Newman1}, the probability that an infected node
$i$ makes infectious contact with an adjacent node $j$ is given by
$T_{ij}=1-\exp(-r_{i}\beta_{ij})$, where $\beta_{ij}$ is the rate of
infectious contact from $i$ to $j$ and $r_{i}$ is the time that $i$ remains
infectious. \ (In this paper, we use \textit{infectious contact} to mean a
contact that results in infection if and only if the recipient is
susceptible.) \ The recovery period $r_{i}$ is a random variable with the
cumulative distribution function (cdf) $F(r)$ and the infectious contact rate
$\beta_{ij}$ has the cdf $F(\beta)$. \ The infectious periods for all
individuals are independent and identically distributed (iid) and the
infectious contact rates for all ordered pairs of individuals are iid. \ \ 

This model can be analyzed by mapping the SIR model onto a semi-directed
network that we call the \textit{epidemic percolation network }\cite{Kenah}.
\ Since the distribution of recovery periods for all nodes and the joint
distribution of contact rates for all pairs of connected nodes are defined
\textit{a priori}, all relevant transmission probabilities can be determined
by assigning the infectious periods and contact rate pairs before an epidemic
begins. \ Starting from the contact network, a single realization of the
epidemic percolation network can be generated as follows:

\begin{enumerate}
\item Choose a recovery period $r_{i}$ for every node $i$ and choose a contact
rate $\beta_{ij}$ for every ordered pair of connected nodes $i$ and $j$ in the
contact network.

\item For each pair of connected nodes $i$ and $j$ in the contact network,
convert the undirected edge between them to a directed edge from $i$ to $j$
with probability
\[
(1-e^{-r_{i}\beta_{ij}})e^{-r_{j}\beta_{ji}},
\]
to a directed edge from $j$ to $i$ with probability
\[
e^{-r_{i}\beta_{ij}}(1-e^{-r_{j}\beta_{ji}}),
\]
and erase the edge completely with probability $\exp(-r_{i}\beta_{ij}%
-r_{j}\beta_{ji})$. \ The edge remains undirected with probability
\[
(1-e^{-r_{i}\beta_{ij}})(1-e^{-r_{j}\beta_{ji}}).
\]

\end{enumerate}

The epidemic percolation network is a semi-directed network that represents a
single realization of the infectious contact process for each connected pair
of nodes, so $4^{m}$ possible epidemic percolation networks exist for a
contact network with $m$ edges. \ The probability of each network is
determined by the underlying SIR model. \ The epidemic percolation network is
very similar to the locally dependent random graph defined by Kuulasmaa
\cite{Kuulasmaa1} for an epidemic on a $d$-dimensional lattice, with two
important differences: First, the underlying structure of the contact network
is not assumed to be a lattice. \ Second, we replace pairs of (occupied)
directed edges between two nodes with a single undirected edge. \ The idea of
the epidemic percolation network is also similar to the idea of forward and
backward branching processes, which have been used to derive the probability
and final size of an epidemic, respectively, in SIR models with independent
infectiousness and susceptibility \cite{Diekmann}. \ The epidemic percolation
network can be thought of as a simultaneous mapping of the forward and
backward branching processes that generalizes to models with arbitrary joint
distributions of infectiousness and susceptibility. \ (The relationship
between epidemic percolation networks and branching processes is discussed
further in Section 5.1.) \ 

In the Appendix, we define epidemic percolation networks for a very general
time-homogeneous stochastic SIR epidemic model (which includes network-based
models and models with random and proportionate mixing as special cases) and
prove that the size distribution of outbreaks starting from node $i$ is
identical to the distribution of its out-component sizes in the corresponding
probability space of percolation networks. \ Because of this equality of
distribution, epidemic percolation networks can be used to analyze a much more
general class of epidemic models than that defined in the Introduction. \ In
this paper, we show how they can be used to analyze stochastic SIR epidemic
models with random or proportionate mixing.

\subsection{Structure of semi-directed networks}

In this subsection, we review the structure of directed and semi-directed
networks as discussed in \cite{Boguna,Meyers,Broder,Dorogovtsev}. \ Reviews of
the structure and analysis of undirected and purely directed networks can be
found in \cite{Albert,Newman2,Newman3,NBW}.

The \textit{indegree} and \textit{outdegree} of node $i$ are the number of
incoming and outgoing directed edges incident to $i$. \ Since each directed
edge is an outgoing edge for one node and an incoming edge for another node,
the mean indegree and outdegree of a semi-directed network are equal. \ The
\textit{undirected degree} of node $i$ is the number of undirected edges
incident to $i$. \ 

A \textit{component} is a maximal group of connected nodes. \ The
\textit{size} of a component is the number of nodes it contains and its
\textit{relative size} is its size divided by the total size of the network.
\ There are four types of components in a semi-directed network. \ 

The \textit{out-component} of node $i$ includes $i$ and all nodes that can be
reached from $i$ by following a series of edges in the proper direction
(undirected edges are bidirectional). \ The \textit{in-component} of node $i$
includes $i$ and all nodes from which $i$ can be reached by following a series
of edges in the proper direction. \ By definition, node $i$ is in the
in-component of node $j$ if and only if $j$ is in the out-component of $i$.
\ Therefore, the mean size of in- and out-components in any semi-directed
network must be equal. \ 

The \textit{strongly-connected component} of a node $i$ is the intersection of
its in- and out-components; it is the set of all nodes that can be reached
from node $i$ and from which node $i$ can be reached. \ All nodes in a
strongly-connected component have the same in-component and the same
out-component. \ The \textit{weakly-connected component} of node $i$ is the
set of nodes that are connected to $i$ when the direction of the edges is
ignored. \ 

\begin{figure}
[ptb]
\begin{center}
\includegraphics[
height=4.1684in,
width=5.5486in
]%
{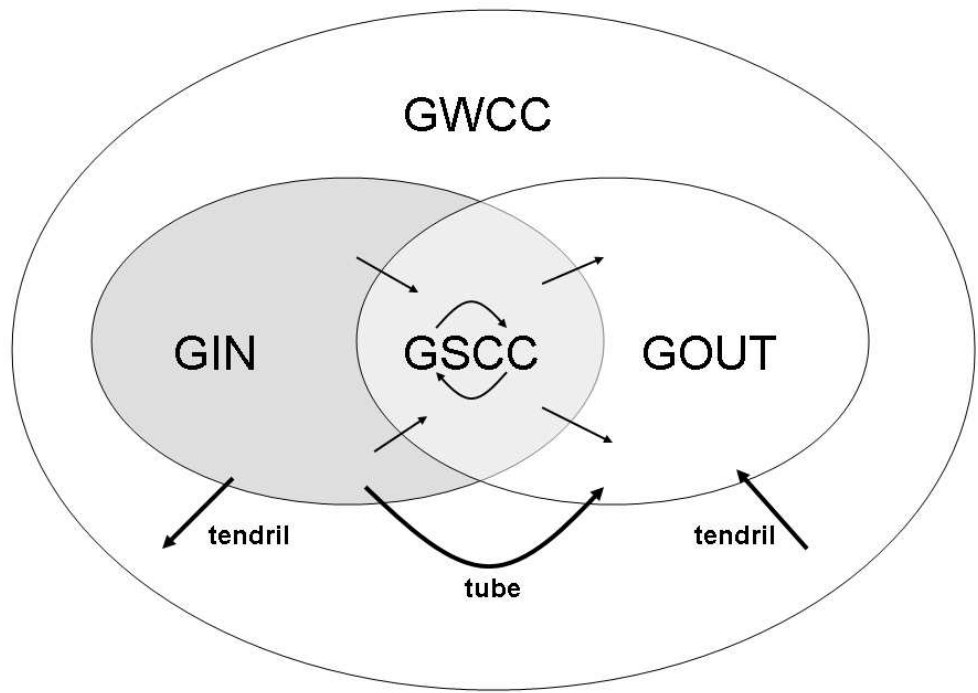}%
\caption{\textquotedblleft Bowtie" diagram showing the giant components,
tendrils, and tubes of a supercritical semi-directed network. \ Adapted from
Broder \textit{et al.} \cite{Broder} and Dorogovtsev \textit{et al.}
\cite{Dorogovtsev}.}%
\label{bowtie}%
\end{center}
\end{figure}

For giant components, we use the definitions given in
\cite{Dorogovtsev,Schwartz}. \ Giant components are so called because they
have asymptotically positive relative size in the limit of a large population.
\ All other components are "small" in the sense that they have asymptotically
zero relative size. \ There are two phase transitions in a semi-directed
network: One where a unique giant weakly-connected component (GWCC) emerges
and another where unique giant in-, out-, and strongly-connected components
(GIN, GOUT, and GSCC) emerge. \ The GWCC contains the other three giant
components. \ The GSCC is the intersection of the GIN and the GOUT, which are
the common in- and out-components of nodes in the GSCC. \ \textit{Tendrils}
are components in the GWCC that are outside the GIN and the GOUT.
\ \textit{Tubes} are directed paths from the GIN to the GOUT that do not
intersect the GSCC. \ All tendrils and tubes are small components. \ A
schematic representation of these components is shown in Figure \ref{bowtie}.

\subsection{Epidemic percolation networks and epidemics}

An \textit{outbreak} begins when one or more nodes are infected from outside
the population. \ These are called \textit{imported infections}. \ The
\textit{final size} of an outbreak is the number of nodes that are infected
before the end of transmission, and its \textit{relative final size} is its
final size divided by the total size of the network. \ The nodes infected in
the outbreak can be identified with the nodes in the out-components of the
imported infections. \ This identification is made mathematically precise in
the Appendix. \ 

We define a \textit{self-limited outbreak} to be an outbreak whose relative
final size approaches zero in the limit of a large population. \ An
\textit{epidemic} is an outbreak whose relative final size is positive in the
limit of a large population. \ For many SIR epidemic models (including the one
in the Introduction), there is an \textit{epidemic threshold:} The probability
of an epidemic is zero below the epidemic threshold and the probability and
relative final size of an epidemic are positive above the epidemic threshold
\cite{Newman1,Andersson,Diekmann,Sander}. \ 

If all out-components in the epidemic percolation network are small, then only
self-limited outbreaks are possible. \ If the epidemic percolation network
contains a GSCC, then any infection in the GIN will lead to the infection of
the entire GOUT. \ Therefore, the epidemic threshold corresponds to the
emergence of the GSCC in the epidemic percolation network. \ For any set of
imported infections, the probability of an epidemic is equal to the
probability that at least one imported infection occurs in the GIN. \ For any
finite set of imported infections, the relative final size of an epidemic is
asymptotically equal to the proportion of the network contained in the GOUT.
\ Although some nodes outside the GOUT may be infected (e.g. tendrils and
tubes), they will constitute a finite number of small components whose total
relative size is asymptotically zero.

This argument can be extended to epidemic percolation networks for
heterogeneous populations. \ The size distribution of outbreaks starting from
an initial infection in any given node $i$ is equal to the distribution of the
out-component sizes of node $i$ in the probability space of epidemic
percolation networks. \ In the limit of a large population, the probability
that the infection of node $i$ causes an epidemic is equal to the probability
that $i$ is in the GIN and the probability that $i$ is infected in an epidemic
is equal to the probability that $i$ is in the GOUT. \ Note that the size
distribution of outbreaks and the probability of an epidemic can depend on the
initial infection(s), but the relative final size of an epidemic does not. \ 

\subsection{Random and proportionate mixing}

In this paper, we show how epidemic percolation networks can be used to
analyze stochastic SIR epidemic models with random or proportionate mixing.
\ Methods exist to calculate the final size distribution of epidemics for such
models in a population of size $n$, but they require solving a recursive
system of $n$ equations \cite{Andersson,Gani}. \ We will show how the size
distribution of outbreaks, the epidemic threshold, and the probability and
relative final size of a large epidemic can be calculated in the limit of
large $n$ by solving a much simpler set of equations. \ These methods also
generalize more easily to heterogeneous populations. \ We will show that these
methods are equivalent to branching processes when the indegree and outdegree
of the epidemic percolation network are independent, and we will use them to
re-derive classical results from several areas of theoretical infectious
disease epidemiology. \ 

The rest of the paper is organized as follows: In Section 2, we find the
degree distributions of the epidemic percolation networks corresponding to SIR
models with random and proportionate mixing. \ In Section 3, we review the use
of probability generating functions to analyze semi-directed networks and show
how these simplify in the case of purely directed networks. \ In Section 4, we
present a series of simulations to show that epidemic percolation networks
accurately predict the mean outbreak size and the probability and final size
of an epidemic for SIR models with random and proportionate mixing. \ In
Section 5, we show that epidemic percolation networks with independent
indegree and outdegree are equivalent to (forward and backward) branching
processes and re-derive classical results from the epidemiology of sexually
transmitted diseases, vector-borne diseases, and controlled diseases. \ In the
Appendix, we show that an epidemic percolation network can be defined for any
time-homogeneous stochastic SIR epidemic model in a closed population and
prove that the distribution of outbreaks starting from a node $i$ is equal to
the distribution of its out-component sizes in the corresponding probability
space of epidemic percolation networks. \ We conclude that the theory of
percolation on semi-directed networks provides a very general framework for
the analysis of stochastic SIR epidemic models in closed populations. \ \ 

\section{Epidemics with random mixing}

In this section, we derive probability generating functions for the degree
distributions of epidemic percolation networks corresponding to SIR models
with random mixing and proportionate mixing. \ The most important difference
between epidemic models with random mixing and network-based models is that
the infectious contact rate between any pair of nodes is inversely
proportional to the population size. \ \cite{Andersson,Diekmann,AndersonMay}.
\ We deal first with random mixing and introduce proportionate mixing as a
generalization. \ To introduce random mixing, we modify the epidemic model
from the Introduction in three ways:

\begin{enumerate}
\item The contact network is always a complete graph, so infection can be
transmitted between any two individuals. \ 

\item We relax that assumption (from \cite{Newman1,Kenah}) that $\beta_{ij}$
and $\beta_{ji}$ are iid. \ Instead, we let $\beta_{ij}$ and $\beta_{ji}$ have
a joint distribution $F(\beta_{ij},\beta_{ji})$ that is symmetric in its
arguments (i.e. $F(\beta_{1},\beta_{2})=F(\beta_{2},\beta_{1})$ for all
$\beta_{1},\beta_{2}$). \ This symmetry forces the joint distribution of
contact rates between any two individuals to be independent of the indices
assigned to them.

\item In a population of size $n$, the contact rate from $i$ to $j$
$\beta_{ij}(n-1)^{-1}$ and the contact rate from $j$ to $i$ is $\beta
_{ji}(n-1)^{-1}$, where $\beta_{ij}$ and $\beta_{ji}$ have the joint
distribution $F(\beta_{ij},\beta_{ji})$ from above.
\end{enumerate}

The epidemic percolation network for a random mixing model is defined in the
same way as that for a network-based model, except that $\beta_{ij}(n-1)^{-1}$
replaces $\beta_{ij}$. \ Let $g(x,y,u|n,r_{i},\beta_{ij};r_{j},\beta_{ji})$ be
the conditional probability generating function (pgf) for the number of
incoming, outgoing, and undirected edges incident to node $i$ that appear
between $i$ and $j$ in the epidemic percolation network given $n$, $r_{i}$,
$\beta_{ij}$, $r_{j}$, and $\beta_{ji}$. \ Then $g(x,y,u|n,r_{i},\beta
_{ij};r_{j},\beta_{ji})$ is
\begin{align*}
& \exp(-\frac{r_{i}\beta_{ij}+r_{j}\beta_{ji}}{n-1})+\exp(-\frac{r_{i}%
\beta_{ij}}{n-1})[1-\exp(-\frac{r_{j}\beta_{ji}}{n-1})]x\\
& +[1-\exp(-\frac{r_{i}\beta_{ij}}{n-1})]\exp(-\frac{r_{j}\beta_{ji}}%
{n-1})y+[1-\exp(-\frac{r_{i}\beta_{ij}}{n-1})][1-\exp(-\frac{r_{j}\beta_{ji}%
}{n-1})]u.
\end{align*}
In the limit of large $n$,
\[
g(x,y,u|n,r_{i},\beta_{ij};r_{j},\beta_{ji})=1-\frac{r_{i}\beta_{ij}%
+r_{j}\beta_{ji}}{n-1}+\frac{r_{j}\beta_{ji}}{n-1}x+\frac{r_{i}\beta_{ij}%
}{n-1}y+o(n^{-1}),
\]
so undirected edges disappear in the limit of a large population. \ Given
$r_{i}$ and $n$, the conditional pgf for the number of incoming, outgoing, and
undirected edges incident to $i$ that appear between $i$ and a node $j\neq i$
can be found by integrating over the distribution of possible $(\beta
_{ij},\beta_{ji})$ and $r_{j}$:
\begin{align*}
g(x,y,u|n,r_{i})  & =\int_{0}^{\infty}\int_{0}^{\infty}\int_{0}^{\infty
}g(x,y,u|\beta,r_{i},r_{j},n)dF(\beta_{ij},\beta_{ji})dF(r_{j})\\
& =1+\frac{E[r]E[\beta](x-1)+r_{i}E[\beta](y-1)}{n-1}+o(n^{-1}).
\end{align*}
Given $r_{i}$ and $n$, the conditional pgf for the total number of incoming
and outgoing edges incident to $i$ in the percolation network is
\[
(1+\frac{E[r]E[\beta](x-1)+r_{i}E[\beta](y-1)}{n-1}+o(n^{-1}))^{n-1}.
\]
In the limit of large $n$, this converges to
\[
G(x,y,u|r_{i})=e^{E[r]E[\beta](x-1)+r_{i}E[\beta](y-1)}.
\]
Therefore, the pgf for the degree distribution of the percolation network in
the limit of a large population is
\begin{equation}
G(x,y,u)=e^{E[r]E[\beta](x-1)}\int_{0}^{\infty}e^{rE[\beta](y-1)}%
dF(r).\label{Gxy}%
\end{equation}

Several results follow immediately from inspection of this function: First,
undirected edges vanish in the limit of a large population, leaving a purely
directed epidemic percolation network. \ Second, the indegree and outdegree of
nodes in the epidemic percolation network are independent. \ Third, the
indegree has a Poisson distribution with mean $E[r]E[\beta]$. \ Finally, the
outdegree has a conditional Poisson distribution for any given recovery period
$r$. \ The mean outdegree is $E[r]E[\beta]$ as required, but the outdegree
distribution is not necessarily Poisson. \ For example, if $r\sim$
exponential($\lambda$), then the outdegree has a geometric($\frac{\lambda
}{\lambda+E[\beta]}$) distribution. \ More generally, if $r\sim$
gamma($\alpha,\lambda$), then the outdegree has a negative binomial($\alpha
,\frac{\lambda}{\lambda+E[\beta]}$) distribution. \ 

\subsection{Proportionate mixing}

A useful generalization of the SIR model with random mixing is to allow the
population to be composed of $K$ distinct subpopulations, where each
subpopulation $k$ constitutes a proportion $w_{k}$ of the overall population.
\ Let $F_{k}(r)$ be the cumulative distribution function for the recovery
period of nodes in subpopulation $k$. \ In addition, let each subpopulation
$k$ have a relative infectiousness $\alpha_{k}$ and a relative susceptibility
$\gamma_{k}$. \ If nodes $i$ and $j$ are in subpopulations $k_{i}$ and $k_{j}%
$, then the infectious contact rate from $i$ to $j$ is $\alpha_{k_{i}}%
\beta_{ij}\gamma_{k_{j}}$ and the infectious contact rate from $j$ to $i$ is
$\alpha_{k_{j}}\beta_{ji}\gamma_{k_{i}}$, where $\beta_{ij}$ and $\beta_{ji}$
have a joint distribution function $F(\beta_{ij},\beta_{ji})$ as before.
\ This formulation for an epidemic model with a heterogeneous population is
called \textit{proportionate mixing }\cite{Andersson,Diekmann}. \ Since the
relative infectiousness and relative susceptibility are each determined only
up to a multiplicative constant, we assume without loss of generality that
\[
\sum_{k=1}^{K}w_{k}E[r_{k}]\alpha_{k}=\sum_{k=1}^{K}w_{k}\gamma_{k}=1.
\]

Let $g_{k_{i}k_{j}}(x,y,u|n,r_{i},\beta_{ij};r_{j},\beta_{ji})$ be the
conditional pgf for the number of incoming, outgoing, and undirected edges
incident to $i$ that appear between nodes $i$ and $j$ in subpopulations
$k_{i}$ and $k_{j}$ in the percolation network given $\beta_{ij}$, $\beta
_{ji}$, $r_{i}$, $r_{j}$, and $n$. \ Then $g_{k_{i}k_{j}}(x,y,u|n,r_{i}%
,\beta_{ij};r_{j},\beta_{ji})$ equals
\[
1-\frac{r_{i}\alpha_{k_{i}}\beta_{ij}\gamma_{k_{j}}+r_{j}\alpha_{k_{j}}%
\beta_{ji}\gamma_{k_{i}}}{n-1}+\frac{r_{j}\alpha_{k_{j}}\beta_{ji}%
\gamma_{k_{i}}}{n-1}x+\frac{r_{i}\alpha_{k_{i}}\beta_{ij}\gamma_{k_{j}}}%
{n-1}y+o(n^{-1}).
\]
Let $g_{k_{i}}(x,y,u|r_{i},n)$ be the conditional pgf for the number of
incoming, outgoing, and undirected edges incident to $i$ that appear between
$i$ and a node $j\neq i$ given $k_{i}$, $r_{i}$, and $n$. \ Then%

\begin{align*}
g_{k_{i}}(x,y,u|r_{i},n)  & =\sum_{k_{j}=1}^{K}\int_{0}^{\infty}\int
_{0}^{\infty}\int_{0}^{\infty}w_{k_{j}}g_{k_{i}k_{j}}(x,y,u|\beta_{ij}%
,\beta_{ji},r_{i},r_{j},n)dF(\beta_{ij},\beta_{ji})dF_{k_{j}}(r_{j})\\
& =\sum_{k_{j}=1}^{K}w_{k_{j}}[1+\frac{E[r_{k_{j}}]\alpha_{k_{j}}%
E[\beta]\gamma_{k_{i}}(x-1)+r_{i}\alpha_{k_{i}}E[\beta]\gamma_{k_{j}}%
(y-1)}{n-1}+o(n^{-1})].
\end{align*}
The conditional pgf for the total number of incoming, outgoing, and undirected
edges incident to node $i$ given $r_{i}$ and $n$ is
\[
\prod_{k=1}^{K}[1+\frac{E[r_{k}]\alpha_{k}E[\beta]\gamma_{k_{i}}%
(x-1)+r_{i}\alpha_{k_{i}}E[\beta]\gamma_{k}(y-1)}{n-1}+o(n^{-1})]^{w_{k}(n-1)}%
\]
In the limit of large $n$, this becomes
\begin{align*}
g_{k_{i}}(x,y,u|r_{i})  & =\prod_{k=1}^{K}e^{w_{k}[E[r_{k}]\alpha_{k}%
E[\beta]\gamma_{k_{i}}(x-1)+r_{i}\alpha_{k_{i}}E[\beta]\gamma_{k}(y-1)]}\\
& =e^{E[\beta]\gamma_{k_{i}}(x-1)+r_{i}\alpha_{k_{i}}E[\beta](y-1)}.
\end{align*}
Integrating over the distribution of infectious periods in subpopulation
$k_{i}$ yields
\begin{align*}
G_{k_{i}}(x,y,u)  & =\int_{0}^{\infty}g_{k_{i}}(x,y,u|r_{i})dF_{k_{i}}%
(r_{i})\\
& =e^{E[\beta]\gamma_{k_{i}}(x-1)}\int_{0}^{\infty}e^{r_{i}\alpha_{k_{i}%
}E[\beta](y-1)}dF_{i_{k}}(r_{i}).
\end{align*}
Note that the indegree distribution of subpopulation $k_{i}$ has a Poisson
distribution, the outdegree distribution is a mixture of Poisson
distributions, and the indegree and outdegree are independent within
subpopulation $k_{i}$. \ Finally, the pgf for the degree distribution of the
epidemic percolation network in the limit of a large population is:
\[
G(x,y,u)=\sum_{k=1}^{K}w_{k}G_{k}(x,y,u).
\]

The proportional mixing assumption has two important consequences that
enormously simplify the analysis of the epidemic percolation network: First,
the probability that an edge terminates at a node in subpopulation $k$ is
proportional to the expected indegree of subpopulation $k$ and independent of
the node at which the edge began. \ Second, the probability that an edge
originates at a node in subpopulation $k$ is proportional to the expected
outdegree of subpopulation $k$ and independent of the node at which the edge
terminates. \ To prove the first, we observe that the total number of directed
edges to nodes in subpopulation $k_{0}$ from a node $i$ in subpopulation
$k_{i}$ with recovery period $r_{i}$ is a sum of $w_{k_{0}}n$ iid Bernoulli
random variables with mean
\[
\frac{r_{i}\alpha_{k_{i}}E[\beta]\gamma_{k_{0}}}{n-1}+o(n^{-1})
\]
In the limit of large $n$, the number of outgoing edges from node $i$ to nodes
in subpopulation $k_{0}$ has a Poisson distribution with mean
\[
w_{k_{0}}(r_{i}\alpha_{k_{i}}E[\beta]\gamma_{k_{0}}).
\]
The total number of outgoing edges from node $i$ has a Poisson distribution
with mean $r_{i}\alpha_{k_{i}}E[\beta]$, which is a sum of $K$ Poisson random
variables with means $w_{k}(r_{i}\alpha_{k_{i}}E[\beta]\gamma_{k})$,
$k=1,...,K$. \ By the strong law of large numbers, the proportion of these
edges that terminate at nodes in subpopulation $k_{0}$ converges almost surely
to $w_{k_{0}}\gamma_{k_{0}}$, which is proportional to the expected indegree
of subpopulation $k_{0}$ and independent of $k_{i}$ and $r_{i}$. \ A similar
argument shows that the proportion of incoming edges to node $i$ that
originate at nodes in subpopulation $k_{0}$ converges almost surely to
$w_{k_{0}}E[r_{k_{0}}]\alpha_{k_{0}}$, which is proportional to the expected
outdegree of subpopulation $k_{0}$ and independent of $k_{i}$ and $r_{i}$.

\section{Components of epidemic percolation networks}

Methods of calculating the size distribution of small components, the
percolation threshold, and the proportion of a network contained in the GIN,
the GOUT, and the GSCC for semi-directed networks with arbitrary degree
distributions have been developed by Bogu\~{n}\'{a} and Serrano \cite{Boguna}%
\ and Meyers, Newman, and Pourbohloul \cite{Meyers}. \ For purely directed and
purely undirected networks, these methods simplify to equations derived by
Newman, Strogatz, and Watts \cite{NSW,Albert,Newman2,Newman3}. \ In this
section, we outline the methods for semi-directed networks and show how they
simplify in the case of purely directed networks. \ This discussion is adapted
from our previous paper \cite{Kenah} and introduces notation that will be used
in the rest of this paper. \ For readers who desire an introduction to random
graphs and percolation on networks, we recommend Albert and Barab\'{a}si
\cite{Albert} and Newman \cite{Newman3}.

The networks considered here have no small loops and no two-point degree
correlations (i.e. the degree of a node reached by following an edge forward
or backward is independent of the degree of the node from which we start).
\ As shown above, this is sufficient for models with random and proportionate
mixing. \ Since these methods assume no clustering of contacts, they do not
apply to epidemics on networks with spatial structure \cite{Kuulasmaa1,Sander}%
, small-world networks \cite{Watts}, or other clustered networks
\cite{BallNeal,Keeling}. \ The development of methods for clustered networks
is an area of active research \cite{Serrano}. \ Nonetheless, the isomorphism
to an epidemic percolation network is valid for any time-homogeneous SIR
model, including models that cannot be analyzed via the generating function
formalism outlined here.

If $a$, $b$, and $c$ are nonnegative integers, let $G^{(a,b,c)}(x,y,u)$ be the
derivative obtained after differentiating $a$ times with respect to $x$, $b$
times with respect to $y$, and $c$ times with respect to $u$. \ Then the mean
indegree of the epidemic percolation network is $G^{(1,0,0)}(1,1,1)$ and the
mean outdegree is $G^{(0,1,0)}(1,1,1)$. \ Let $\left\langle k_{d}\right\rangle
$ denote the common mean of the directed degrees. \ The mean undirected degree
is $\left\langle k_{u}\right\rangle =G^{(0,0,1)}(1,1,1)$. \ For the epidemic
percolation network for the homogeneous SIR model with random mixing,
$\left\langle k_{d}\right\rangle =$ $E[r]E[\beta]$ and $\left\langle
k_{u}\right\rangle =0$. \ 

Let $G_{f}(x,y,u)$ be the pgf for the degree distribution of a node reached by
going forward along a directed edge, excluding the edge used to reach the
node. \ Since the probability of reaching any node by following a directed
edge is proportional to its indegree,
\[
G_{f}(x,y,u)=\frac{1}{\left\langle k_{d}\right\rangle }\sum_{j,k,m}%
jp_{jkm}x^{j-1}y^{k}u^{m}=\frac{1}{\left\langle k_{d}\right\rangle
}G^{(1,0,0)}(x,y,u).
\]
Similarly, the pgf for the degree distribution of a node reached by going in
reverse along a directed edge (excluding the edge used to reach it) is
\[
G_{r}(x,y,u)=\frac{1}{\left\langle k_{d}\right\rangle }G^{(0,1,0)}(x,y,u).
\]
and the pgf for the degree distribution of a node reached by following an
undirected edge (excluding the edge used to reach it) is
\[
G_{u}(x,y,u)=\frac{1}{\left\langle k_{u}\right\rangle }G^{(0,0,1)}(x,y,u).
\]
The above definitions require that $\left\langle k_{d}\right\rangle >0$ and
$\left\langle k_{u}\right\rangle >0$. \ In a purely undirected network (i.e.
$\left\langle k_{d}\right\rangle =0$), we arbitrarily set $G_{f}%
(x,y,u)=G_{r}(x,y,u)=1$ for all $x$, $y$, and $u$. $\ $In a purely directed
network (i.e. $\left\langle k_{u}\right\rangle =0$), we arbitrarily set
$G_{u}(x,y,u)=1$ for all $x$, $y$, and $u$.

\subsection{Out-components}

Let $H_{f}^{out}(z)$ be the pgf for the size of the out-component at the end
of a directed edge and $H_{u}^{out}(z)$ be the pgf for the size of the
out-component at the "end" of an undirected edge. \ Then, in the limit of a
large population,
\begin{subequations}
\begin{align}
H_{f}^{out}(z)  & =zG_{f}(1,H_{f}^{out}(z),H_{u}^{out}(z)),\label{Hf,out}\\
H_{u}^{out}(z)  & =zG_{u}(1,H_{f}^{out}(z),H_{u}^{out}(z)).\label{Hu,out}%
\end{align}
The pgf for the out-component size of a randomly chosen node is
\end{subequations}
\begin{equation}
H^{out}(z)=zG(1,H_{f}^{out}(z),H_{u}^{out}(z)).\label{Hout}%
\end{equation}
In a purely directed network, $H_{u}^{out}(z)=1$ for all $z$ because
$G_{u}(x,y,u)=1$ for all $x$, $y$, and $u$. $\ $Thus $H_{f}^{out}%
(z)=zG_{f}(1,H_{f}^{out}(z),1)$ and $H^{out}(z)=zG(1,H_{f}^{out}(z),1)$. \ 

Given power series for $H_{f}^{out}(z)$ and $H_{u}^{out}(z)$ that are accurate
to $z^{n}$, equations (\ref{Hf,out}) and (\ref{Hu,out}) can be used to obtain
series that are accurate to $z^{n+1}$. \ With power series for $H_{f}%
^{out}(z)$ and $H_{u}^{out}(z)$ that are accurate to $z^{n}$, equation
(\ref{Hout}) can be used to obtain a power series for $H^{out}(z)$ that is
accurate to $z^{n+1}$. \ The coefficients on $z^{0}$ in $H_{f}^{out}(z)$ and
$H_{u}^{out}(z)$ are $G_{f}(1,0,0)$ and $G_{u}(1,0,0)$ respectively.
\ Therefore, power series for $H_{f}^{out}(z)$, $H_{u}^{out}(z)$, and
$H^{out}(z)$ can be computed to any desired order. \ For any $z\in\lbrack
0,1]$, $H_{f}^{out}(z)$ and $H_{u}^{out}(z)$ can be calculated with arbitrary
precision by iterating equations (\ref{Hf,out}) and (\ref{Hu,out}) starting
from initial values $y_{0},u_{0}\in\lbrack0,1)$. \ Estimates of $H_{f}%
^{out}(z)$ and $H_{u}^{out}(z)$ can be used to obtain estimates of
$H^{out}(z)$ with arbitrary precision. \ 

In the limit of a large population, the probability that a node has a finite
out-component is $H^{out}(1)$, so the probability that a randomly chosen node
is in the GIN is $1-H^{out}(1)$. \ The expected size of the out-component of a
randomly chosen node is $H^{out\prime}(1)$. \ Taking derivatives in equation
(\ref{Hout}) yields
\begin{equation}
H^{out\prime}(1)=1+\left\langle k_{d}\right\rangle H_{f}^{out\prime
}(1)+\left\langle k_{u}\right\rangle H_{u}^{out\prime}(1).\label{Hout'(1)}%
\end{equation}
Taking derivatives in equations (\ref{Hf,out}) and (\ref{Hu,out}) and using
the fact that $H_{f}^{out}(1)=H_{u}^{out}(1)=1$ below the epidemic threshold
yields a set of linear equations for $H_{f}^{out\prime}(1)$ and $H_{u}%
^{out\prime}(1)$. \ These can be solved to yield
\begin{equation}
H_{f}^{out\prime}(1)=\frac{1+G_{f}^{(0,0,1)}-G_{u}^{(0,0,1)}}{(1-G_{f}%
^{(0,1,0)})(1-G_{u}^{(0,0,1)})-G_{f}^{(0,0,1)}G_{u}^{(0,1,0)}}%
,\label{Hfout'(1)}%
\end{equation}
and
\begin{equation}
H_{u}^{out\prime}(1)=\frac{1-G_{f}^{(0,1,0)}+G_{u}^{(0,1,0)}}{(1-G_{f}%
^{(0,1,0)})(1-G_{u}^{(0,0,1)})-G_{f}^{(0,0,1)}G_{u}^{(0,1,0)}}%
,\label{Huout'(1)}%
\end{equation}
where the argument of all derivatives is $(1,1,1)$. \ In a purely directed
network, all derivatives involving $G_{u}$ are zero, so
\[
H_{f}^{out\prime}(1)=\frac{1}{1-G_{f}^{(0,1,0)}}%
\]
and
\[
H^{out\prime}(1)=1+\left\langle k_{d}\right\rangle H_{f}^{out\prime}(1).
\]

\subsection{In-components}

The in-component size distribution of a semi-directed network can be derived
using the same logic used to find the out-component size distribution, except
that we consider going backwards along edges. \ Let $H_{r}^{in}(z)$ be the pgf
for the size of the in-component at the beginning of a directed edge,
$H_{u}^{in}(z)$ be the pgf for the size of the in-component at the "beginning"
of an undirected edge, and $H^{in}(z)$ be the pgf for the in-component size of
a randomly chosen node. \ Then
\begin{subequations}
\label{Hur,in}%
\begin{align}
H_{r}^{in}(z)  & =zG_{r}(H_{r}^{in}(z),1,H_{u}^{in}(z)),\label{Hr,in}\\
H_{u}^{in}(z)  & =zG_{u}(H_{r}^{in}(z),1,H_{u}^{in}(z)),\label{Hu,in}\\
H^{in}(z)  & =zG(H_{r}^{in}(z),1,H_{u}^{in}(z)).\label{Hin}%
\end{align}
Power series to arbitrary degrees and numerical estimates with arbitrary
precision can be obtained for $H_{r}^{in}(z)$, $H_{u}^{in}(z)$, and
$H^{in}(z)$ by iterating these equations in the manner described for
$H_{f}^{out}(z)$, $H_{u}^{out}(z)$, and $H^{out}(z)$. \ In a purely directed
network, $H_{u}^{in}(z)=1$ for all $z$ because $G_{u}(x,y,u)=1$ for all $x$,
$y$, and $u$. $\ $Thus $H_{r}^{in}(z)=zG_{r}(H_{r}^{in}(z),1,1)$ and
$H^{in}(z)=zG(H_{r}^{in}(z),1,1)$. \ 

In the limit of a large population, the probability that a node has a finite
in-component is $H^{in}(1)$, so the probability that a randomly chosen node is
in the GOUT is $1-H^{in}(1)$. \ The expected size of the in-component of a
randomly chosen node is $H^{in\prime}(1)$. \ Taking derivatives in equation
(\ref{Hin}) yields
\end{subequations}
\begin{equation}
H^{in\prime}(1)=1+\left\langle k_{d}\right\rangle H_{r}^{in\prime
}(1)+\left\langle k_{u}\right\rangle H_{u}^{in\prime}(1).\label{Hin'(1)}%
\end{equation}
Taking derivatives in equations (\ref{Hr,in}) and (\ref{Hu,in}) and using the
fact that $H_{r}^{in}(1)=H_{u}^{in}(1)=1$ in a subcritical network yields
\begin{equation}
H_{r}^{in\prime}(1)=\frac{1+G_{r}^{(0,0,1)}-G_{u}^{(0,0,1)}}{(1-G_{r}%
^{(1,0,0)})(1-G_{u}^{(0,0,1)})-G_{r}^{(0,0,1)}G_{u}^{(1,0,0)}}%
,\label{Hrin'(1)}%
\end{equation}
and
\begin{equation}
H_{u}^{in\prime}(1)=\frac{1-G_{r}^{(1,0,0)}+G_{u}^{(1,0,0)}}{(1-G_{r}%
^{(1,0,0)})(1-G_{u}^{(0,0,1)})-G_{r}^{(0,0,1)}G_{u}^{(1,0,0)}}%
,\label{Huin'(1)}%
\end{equation}
where the argument of all derivatives is $(1,1,1)$. \ In a purely directed
network, all derivatives involving $G_{u}$ are zero, so
\[
H_{r}^{in\prime}(1)=\frac{1}{1-G_{r}^{(1,0,0)}}%
\]
and
\[
H^{in\prime}(1)=1+\left\langle k_{d}\right\rangle H_{r}^{in\prime}(1).
\]

\subsection{Epidemic threshold}

The epidemic threshold occurs when the expected size of the in- and
out-components in the network becomes infinite. \ Equations (\ref{Hfout'(1)})
and (\ref{Huout'(1)}) show that the mean out-component size becomes infinite
when
\[
(1-G_{f}^{(0,1,0)})(1-G_{u}^{(0,0,1)})-G_{f}^{(0,0,1)}G_{u}^{(0,1,0)}=0,
\]
and equations (\ref{Hrin'(1)}) and (\ref{Huin'(1)}) show that the mean
in-component size becomes infinite when
\[
(1-G_{r}^{(1,0,0)})(1-G_{u}^{(0,0,1)})-G_{r}^{(0,0,1)}G_{u}^{(1,0,0)}=0.
\]
From the definitions of $G_{f}(x,y,u)$, $G_{r}(x,y,u)$ and $G_{u}(x,y,u)$,
both conditions are equivalent to
\[
(1-\frac{1}{\left\langle k_{d}\right\rangle }G^{(1,1,0)})(1-\frac
{1}{\left\langle k_{u}\right\rangle }G^{(0,0,2)})-\frac{1}{\left\langle
k_{d}\right\rangle \left\langle k_{u}\right\rangle }G^{(1,0,1)}G^{(0,1,1)}=0.
\]
Therefore, there is a single epidemic threshold where the GSCC, the GIN, and
the GOUT appear simultaneously. \ 

In a purely directed network, the condition for this epidemic threshold is
much simpler because all derivatives with respect to $u$ and all derivatives
of $G_{u}$ are zero: The mean out-component size becomes infinite when
$1-G_{f}^{(0,1,0)}=0$ and the mean in-component size becomes infinite when
$1-G_{r}^{(1,0,0)}=0$. \ Both of these conditions are equivalent to
$1-\left\langle k_{d}\right\rangle ^{-1}G^{(1,1,0)}=0$. \ 

\subsection{Giant strongly-connected component}

In the limit of a large population, a node is in the GSCC if and only if its
in- and out-components are both infinite. \ A randomly chosen node has a
finite in-component with probability $G(H_{r}^{in}(1),1,H_{u}^{in}(1))$ and a
finite out-component with probability $G(1,H_{f}^{out}(1),H_{u}^{out}(1))$.
\ The probability that a node reached by following an undirected edge has
finite in- and out-components is the solution to the equation
\[
v=G_{u}(H_{r}^{in}(1),H_{f}^{out}(1),v),
\]
and the probability that a randomly chosen node has finite in- and
out-components is $G(H_{r}^{in}(1),H_{f}^{out}(1),v)$ \cite{Boguna}. \ Thus,
the relative size of the GSCC is
\[
1-G(H_{r}^{in}(1),1,H_{u}^{in}(1))-G(1,H_{f}^{out}(1),H_{u}^{out}%
(1))+G(H_{r}^{in}(1),H_{f}^{out}(1),v).
\]
In a purely directed network, this simplifies to
\[
1-G(H_{r}^{in}(1),1)-G(1,H_{f}^{out}(1))+G(H_{r}^{in}(1),H_{f}^{out}(1)).
\]

\section{Simulations}

The following series of simulations provides some examples of how an epidemic
percolation network can be derived from an\ SIR model with random or
proportionate mixing and used to analyze it. \ There are three series of
homogeneous population models and three series of heterogeneous population
models. \ Predictions for the mean size of outbreaks and the probability and
final size of an epidemic were easily obtained and consistently accurate.
\ Models were run on Berkeley Madonna 8.0.1 (\copyright 1997-2000 Robert I.
Macey \& George F. Oster) and Mathematica 5.0.0.0 (\copyright 1988-2003
Wolfram Research, Inc.). \ 

All simulations began with a single imported infection randomly chosen from
the population. \ Simulations in Berkeley Madonna used a Poisson approximation
to the number of new infections in each time step $dt$, with $dt=.005$.
\ Simulations in Mathematica were based on the general stochastic SIR model
from the Appendix: A recovery time for each infected individual was sampled
from the appropriate distribution. \ When person $i$ was infected, an
infectious contact interval for each ordered pair $ij$, $j\neq i $, was
sampled from the appropriate distribution, and the corresponding infectious
contact times were calculated. \ The minimum infectious contact time for each
susceptible individual was stored. \ The next infection occurred in the
susceptible with the smallest infectious contact time. \ The epidemic ended
when the minimum infectious contact time among the remaining susceptibles was
infinite. \ 

An epidemic was defined to be an outbreak that infected more than 10\% or 15\%
of the population. \ These percentages were chosen to obtain an outbreak size
much larger than the expected size of self-limited outbreaks and much smaller
than the expected size of an epidemic. \ For $R_{0}$ near one, the mean
self-limited outbreak size increases and the expected size of an epidemic
decreases, leading to poor separation between self-limited outbreaks and
epidemics. \ Since the mean size of self-limited outbreaks approaches a
constant and the mean size of an epidemic scales with the population size,
this separation can be restored by taking a larger population size. \ However,
the computational time required for an exact simulation varies roughly with
the square of the population size. \ Thus, we did not attempt simulations for
$R_{0}$ below $1.25$ or $1.5$. \ 

In this section and the remainder of the paper, we deal exclusively with
purely directed epidemic percolation networks. \ To simplify notation, we drop
the variable $u$ from $G(x,y,u)$.

\subsection{Homogeneous populations}

The first series of homogeneous population models had a fixed recovery time,
the second series had exponentially-distributed recovery times, and the third
series had ten different recovery time distributions. \ All models had a
single imported infection randomly chosen from the population, a mean recovery
time of one, and a basic reproductive number of $R_{0}$. \ 

When the recovery time is fixed, the pgf for the degree distribution of the
epidemic percolation network is $G(x,y)=e^{R_{0}(x-1)}e^{R_{0}(y-1)}$, so the
indegree and outdegree have independent Poisson distributions with mean
$R_{0}$. \ The model was run at $R_{0}=1.25$, $1.5$, $2$, $2.5$, $3$, $4$, and
$5$. \ At each $R_{0}$, the model was run $10,000$ times in a population of
$10,000$ individuals. \ 

When the recovery time is exponentially distributed, the pgf for the degree
distribution of the epidemic percolation network is
\[
G(x,y)=\frac{e^{R_{0}(x-1)}}{1-R_{0}(y-1)},
\]
so the indegree has a Poisson distribution and the outdegree has a geometric
distribution. \ The mean indegree and outdegree are both $R_{0}$. \ The model
was run at $R_{0}=1.5$, $2$, $2.5$, $3$, $4$, $5$, $6$, $7$, $8$, $9$, and
$10$. \ At each $R_{0}$, the model was run $10,000$ times with a population of
$10,000$ individuals. \ 

Details of the recovery time distributions for the third series of simulations
are shown in Table \ref{RecDistributions}. \ As in the other homogeneous
population models, the indegree and outdegree are independent. \ The pgf for
the indegree is $e^{R_{0}(x-1)}$. \ The pgf for the outdegree is
\[
G^{out}(y)=\int_{0}^{\infty}e^{R_{0}(y-1)r}dF(r).
\]
The pgf for the degree distribution of the epidemic percolation network is
$G(x,y)=e^{R_{0}(x-1)}G^{out}(y)$. \ For each recovery time distribution,
models were run $2,000$ times with a population of $1,000$ individuals at
$R_{0}=1.5$, $2$, $2.5$, $3$, $4$, and $5$. \ 

\begin{figure}
[ptb]
\begin{center}
\includegraphics[
height=4.7859in,
width=5.5486in
]%
{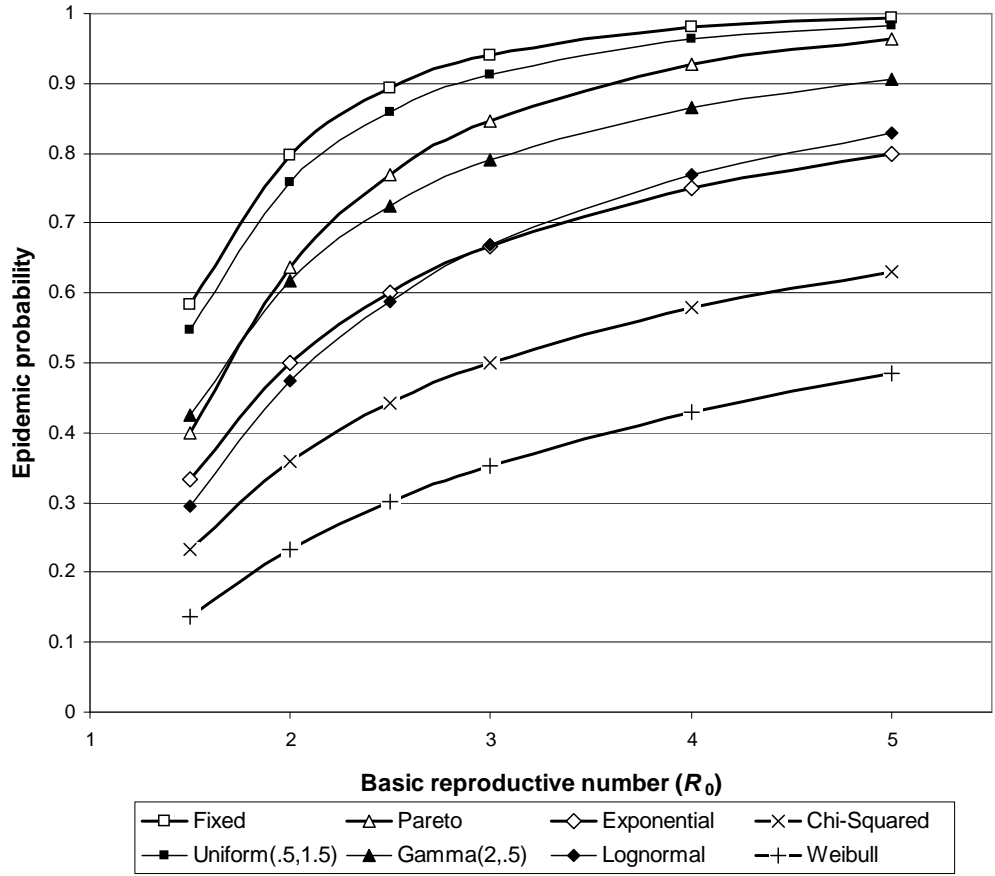}%
\caption{Probability of an epidemic in a homogeneous population as a function
of $R_{0}$ for recovery time distributions from Table \ref{RecDistributions}.}%
\label{P(epi)fig}%
\end{center}
\end{figure}

\begin{figure}
[ptb]
\begin{center}
\includegraphics[
height=5.393in,
width=5.5478in
]%
{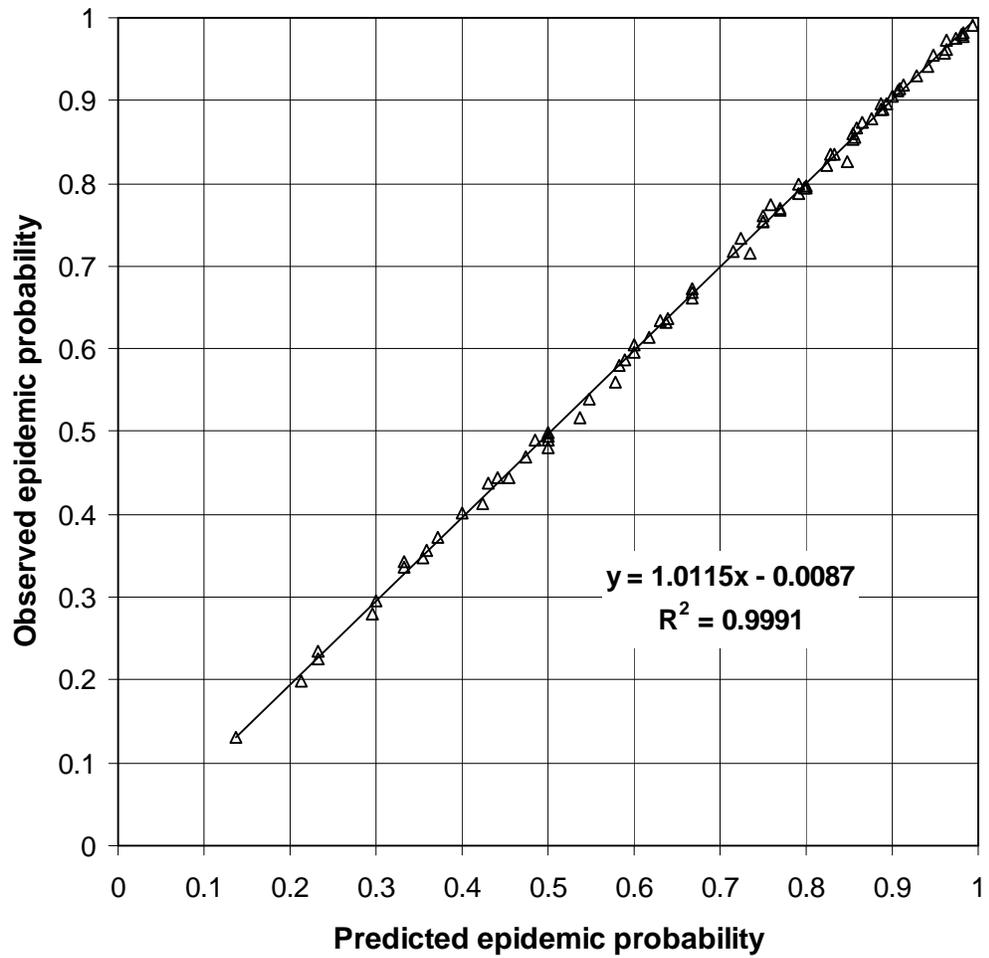}%
\caption{Scatterplot of observed and predicted epidemic probabilities for all
homogeneous population models, with the linear regression equation and $R^{2}%
$. \ \ }%
\label{HomP(epi)}%
\end{center}
\end{figure}

\begin{figure}
[ptb]
\begin{center}
\includegraphics[
height=4.4547in,
width=5.5486in
]%
{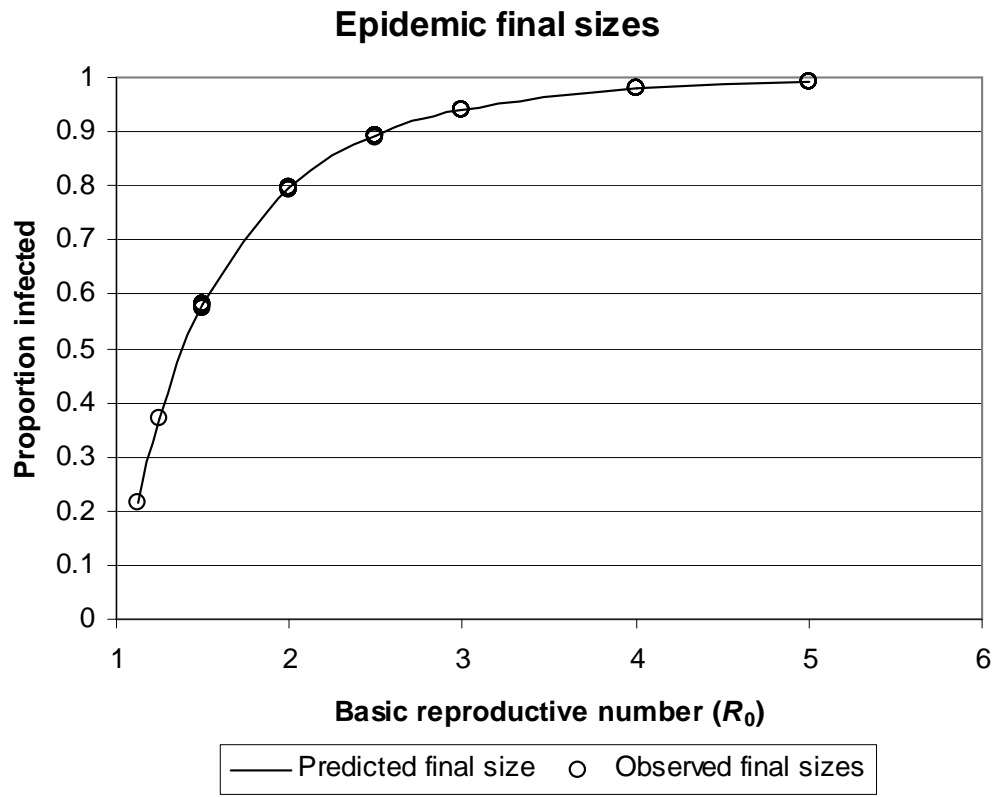}%
\caption{Predicted and observed final sizes of epidemics as a function of
$R_{0}$ for all homogeneous population models. \ The predicted final size was
the same for all recovery time distributions. }%
\label{HomFS}%
\end{center}
\end{figure}

In all homogeneous population models, an epidemic was said to occur when more
than $10\%$ of the population was infected. \ As outlined in Section 3, the
predicted probability of an epidemic is $1-H^{out}(1)$ and the predicted final
size of an epidemic is $1-H^{in}(1)$. \ The predicted final size of an
epidemic for a given $R_{0}$ was the same for all recovery time distributions,
but Figure \ref{P(epi)fig} shows that the probability of an epidemic depends
on both $R_{0}$ and the recovery time distribution. \ The probability and
final size of an epidemic were equal only when the recovery time was fixed.
\ For all other recovery time distributions, the probability of an epidemic
was less than its final size (this inequality is proven in \cite{Kenah}).
\ Figure \ref{HomP(epi)} shows a good agreement between the predicted and
observed probabilities of an epidemic and Figure \ref{HomFS} shows a good
agreement between the observed and predicted final sizes of epidemics. \ 

\subsection{Heterogeneous populations}

In the first two series of heterogeneous population models, the population
consisted of two subpopulations $A$ and $B$ of equal size. \ The average
number of infectious contacts made by a member of subpopulation $B$ during his
or her recovery period is $\lambda$. \ We assumed the following infectious
contact rates from $i$ to $j$ in a population of size $n$: $\frac{8}{3}%
\lambda(n-1)^{-1}$ when $i$ and $j$ are both members of subpopulation $A$,
$\frac{4}{3}\lambda(n-1)^{-1}$ when $i$ and $j$ are members of different
subpopulations, and $\frac{2}{3}\lambda(n-1)^{-1}$ when $i$ and $j $ are both
in subpopulation $B$. \ All models had a mean recovery time of one. \ With
these assumptions, the mean indegree and outdegree of subpopulation $A$ were
$2\lambda$ and the mean indegree and outdegree of subpopulation $B$ were
$\lambda$, producing a positive correlation between susceptibility and infectiousness.

When \ the recovery period is fixed, the pgf for the degree distribution of
the epidemic percolation network is
\[
G(x,y)=.5e^{2\lambda(x-1)}e^{2\lambda(y-1)}+.5e^{\lambda(x-1)}e^{\lambda
(y-1)}.
\]
When the recovery period is exponentially distributed, the pgf for the degree
distribution of the epidemic percolation network is
\[
G(x,y)=.5\frac{e^{2\lambda(x-1)}}{1-2\lambda(y-1)}+.5\frac{e^{\lambda(x-1)}%
}{1-\lambda(y-1)}%
\]
Each model was run at $\lambda=.65,.7,.8,.9,1.0,1.1,1.2,1.3,$ and $1.4$. \ At
each $\lambda$, the model was run $10,000$ times. \ For models with
$\lambda>.65$, the population was $10,000$. \ For $\lambda=.65$, the
population size was $100,000$. \ When $\lambda=.65$, epidemics occur even
though the mean degree of the network is less than one. \ Both series of
models were implemented in Berkeley Madonna.\ 

A third series of simulations was conducted in populations with various
mixtures of recovery time distributions. \ The population had $1,000$
individuals partitioned into subpopulations $A$ and $B$ of $500$ individuals
each. \ Each model was run under two scenarios:\ In the first scenario,
subpopulation $A$ is twice as infectious per unit time and has the same mean
recovery time as subpopulation $B$. \ In the second scenario, both
subpopulations are equally infectious per unit time but the mean recovery
period of subpopulation $A$ is twice as long as that of $B$. \ The degree
distribution of the epidemic percolation network is identical under both
scenarios. \ In the first scenario, models were run for all nine possible
combinations of the Fixed$(1)$, Uniform$(0,2)$, and Exponential$(1)$ recovery
time distributions. \ In the second scenario, models were run for all nine
possible combinations of Fixed$(2)$, Uniform$(0,4)$, and Exponential$(.5)$ in
subpopulation $A$ and Fixed$(1)$, Uniform$(0,2)$, and Exponential$(1)$ in
subpopulation $B$. \ These recovery time distributions are described in Table
\ref{RecDistributions}. \ Every model was run $5,000$ times with $\lambda=2$.
\ An epidemic was defined as an outbreak that infected more than $15\%$ of the
population. \ A similar set of simulations was conducted where subpopulation
$A$ had a mean indegree of $\lambda$ and a mean outdegree of $2\lambda$ while
subpopulation $B$ had a mean indegree of $2\lambda$ and a mean outdegree of
$\lambda$, producing a negative correlation between infectiousness and
susceptibility. \ All of these models were implemented in Mathematica.

\begin{figure}
[ptb]
\begin{center}
\includegraphics[
height=4.5558in,
width=5.5486in
]%
{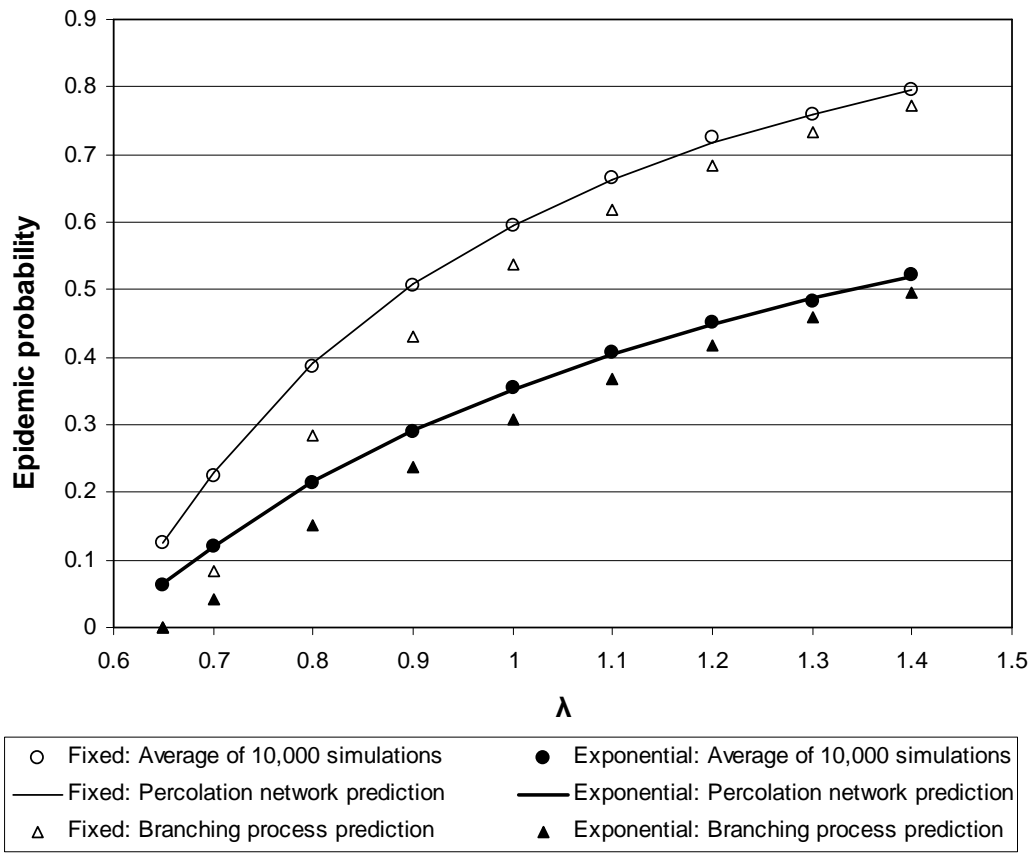}%
\caption{Predicted and observed probabilities of an epidemic as a function of
$\lambda$ for the first two series of heterogeneous population models. \ }%
\label{BranchingPr(epi)}%
\end{center}
\end{figure}

\begin{figure}
[ptb]
\begin{center}
\includegraphics[
height=3.8795in,
width=5.5486in
]%
{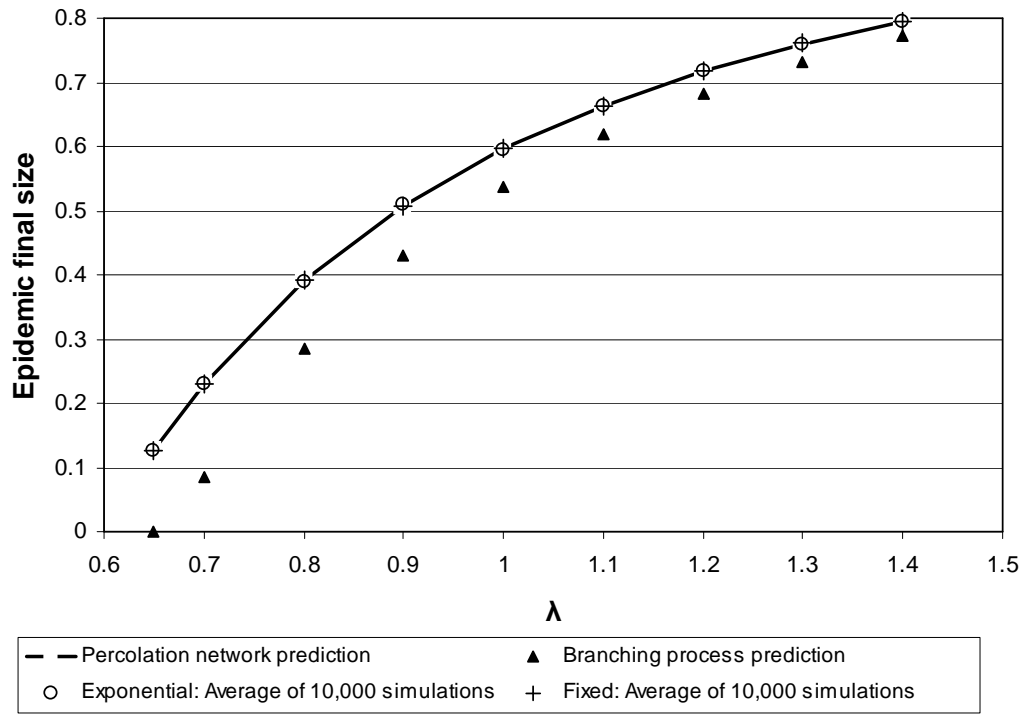}%
\caption{Predicted and observed final sizes of an epidemic as a function of
$\lambda$ for the first two series of heterogeneous population models. \ The
final sizes of epidemics are the same for both fixed and
exponentially-distributed recovery times.}%
\label{BranchingSize}%
\end{center}
\end{figure}

In the first two series of heterogeneous models, an epidemic was defined to
occur when more than $10\%$ of the population was infected. \ The predicted
probability and final size of an epidemic are $1-H^{out}(1)$ and $1-H^{in}(1)
$, respectively, according to the epidemic percolation network. \ According to
a branching process approximation, the predicted probability of an epidemic is
$1-h^{out}(1)$, where
\[
h^{out}(z)=zG(1,h^{out}(z)),
\]
and the predicted final size of an epidemic is $1-h^{in}(1)$, where
\[
h^{in}(z)=zG(h^{in}(z),1).
\]
Figures \ref{BranchingPr(epi)} and \ref{BranchingSize} compare the epidemic
percolation network and branching process predictions of the probability and
final size of an epidemic. \ These models have a positive correlation between
infectiousness and susceptibility, so persons infected through indigenous
transmission are more infectious than persons randomly selected from the
population. \ Since the branching process approximation implicitly assumes
that persons infected through indigenous transmission have the same outdegree
distribution as the general population, it consistently underestimates both
the probability and final size of an epidemic. \ The epidemic percolation
network consistently predicts the correct probability and final size of an
epidemic. \ 

\begin{figure}
[ptb]
\begin{center}
\includegraphics[
height=5.3601in,
width=5.5478in
]%
{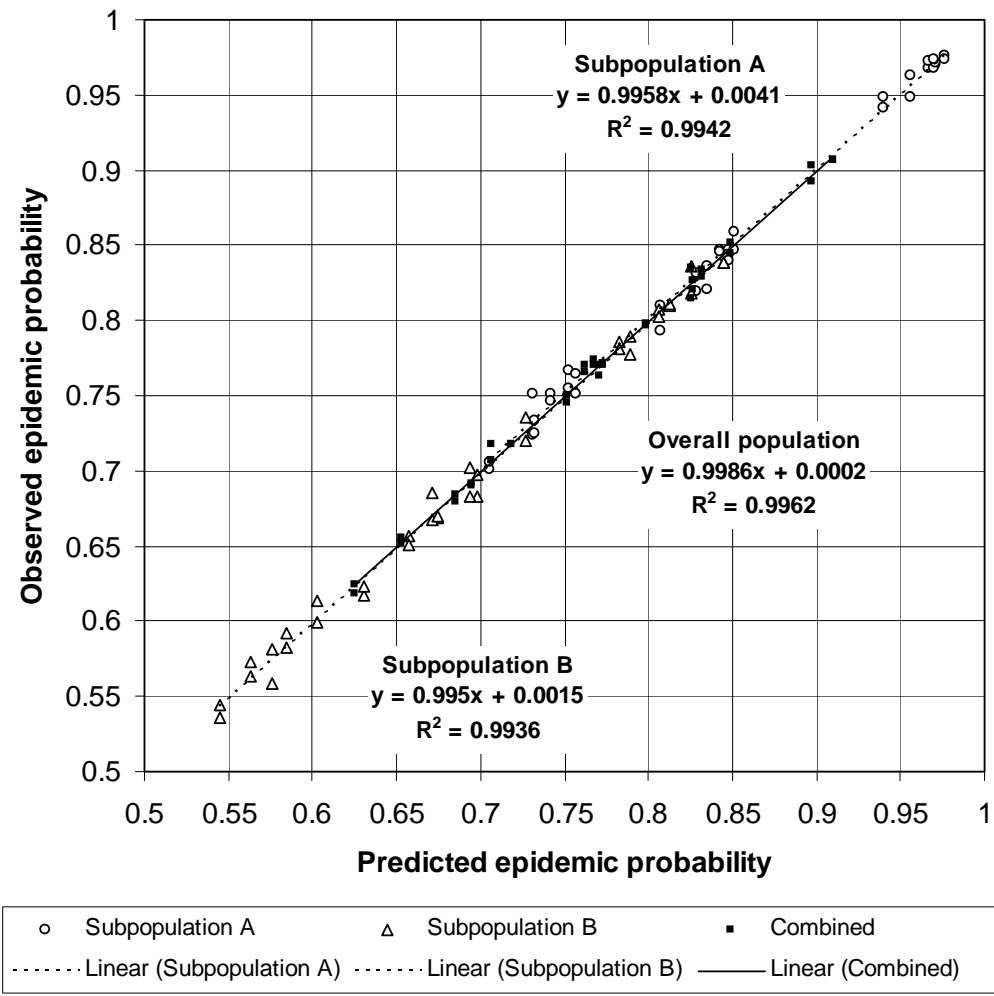}%
\caption{Scatterplot of observed and predicted epidemic probabilities in the
third series of heterogeneous population models. \ The "subpopulation A" and
"subpopulation B" points show conditional epidemic probabilities given an
initial case in subpopulation A and B respectively. \ The "combined" points
show the epidemic probability when the initial case is randomly chosen from
the entire population. \ The linear regression equation and $R^{2}$ are shown
separately for all three sets of points. }%
\label{HetPr(epi)}%
\end{center}
\end{figure}

\begin{figure}
[ptb]
\begin{center}
\includegraphics[
height=5.3938in,
width=5.5478in
]%
{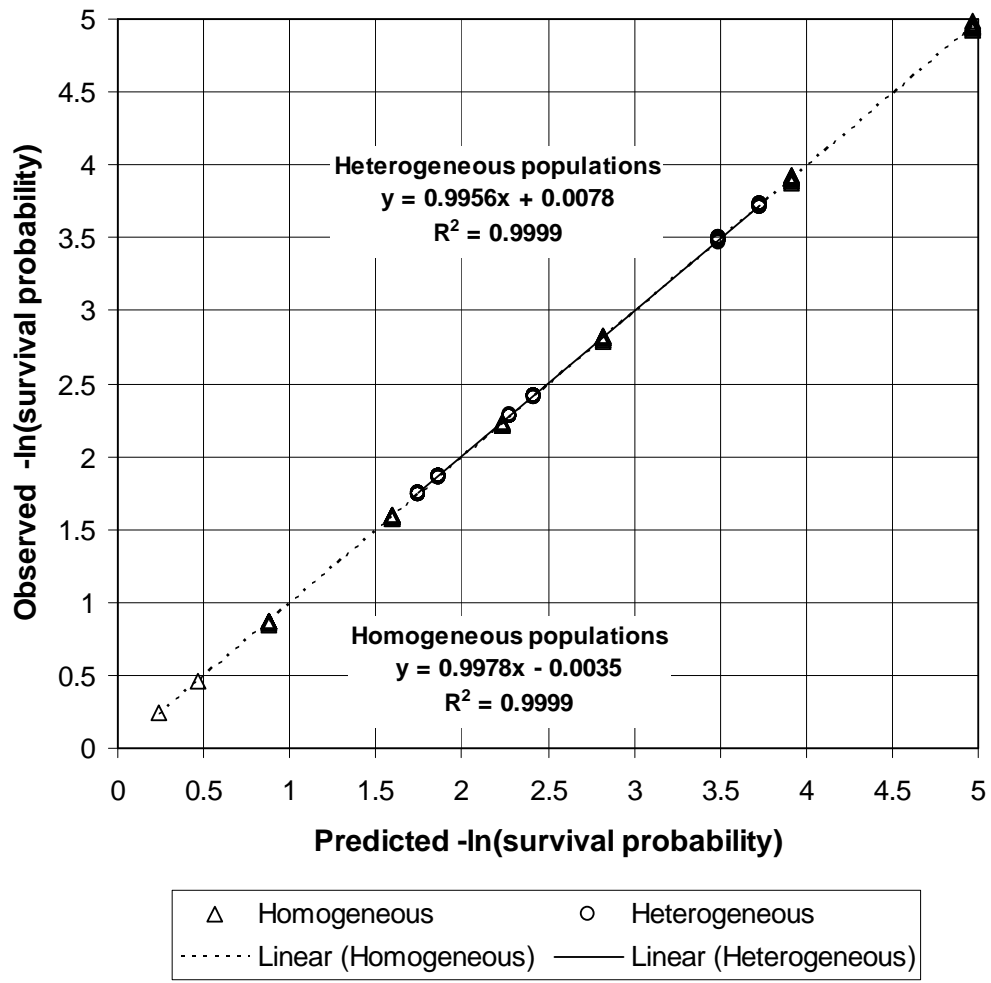}%
\caption{Scatterplot of the observed and predicted cumulative hazard of
infection in an epidemic. \ Linear regression equations and $R^{2}$ are shown
separately for homogeneous and heterogeneous population models.}%
\label{Survival}%
\end{center}
\end{figure}

Figure \ref{HetPr(epi)} shows a scatterplot of observed and predicted epidemic
probabilities for all heterogeneous population models. \ The predicted
probability of an epidemic for an initial case randomly chosen from the
population is $1-H^{out}(1)$. \ The \textquotedblleft combined" points show
the observed and predicted epidemic probabilities for an initial case randomly
chosen from the overall population. \ Let $G_{A}(x,y)$ and $G_{B}(x,y)$ be the
pgf of the degree distributions of nodes in subpopulations $A$ and $B$
respectively. \ The predicted probability of an epidemic for an initial case
chosen randomly from subpopulation $A$ is $1-G_{A}(1,H_{f}^{out}(1))$, and the
predicted probability of an epidemic for an initial case chosen randomly from
subpopulation $B$ is $1-G_{B}(1,H_{f}^{out}(1))$. \ The \textquotedblleft
subpopulation $A$" points show the observed and predicted epidemic
probabilities for an initial case randomly chosen from subpopulation $A$, and
the \textquotedblleft subpopulation $B$" points show the observed and
predicted epidemic probabilities for an initial case randomly chosen from
subpopulation $B$. \ All three sets of points are close to the diagonal,
showing that epidemic percolation networks accurately predicted the
probability of an epidemic. \ The predicted cumulative hazard of infection in
an epidemic is $-\ln(H^{in}(1))$. \ Figure \ref{Survival} shows a scatterplot
of the observed and predicted cumulative hazard of infection in an epidemic
for all heterogeneous population models. \ All points are close to the
diagonal, showing that epidemic percolation networks accurately predicted the
final size of epidemics.

\begin{figure}
[ptb]
\begin{center}
\includegraphics[
height=5.3601in,
width=5.5478in
]%
{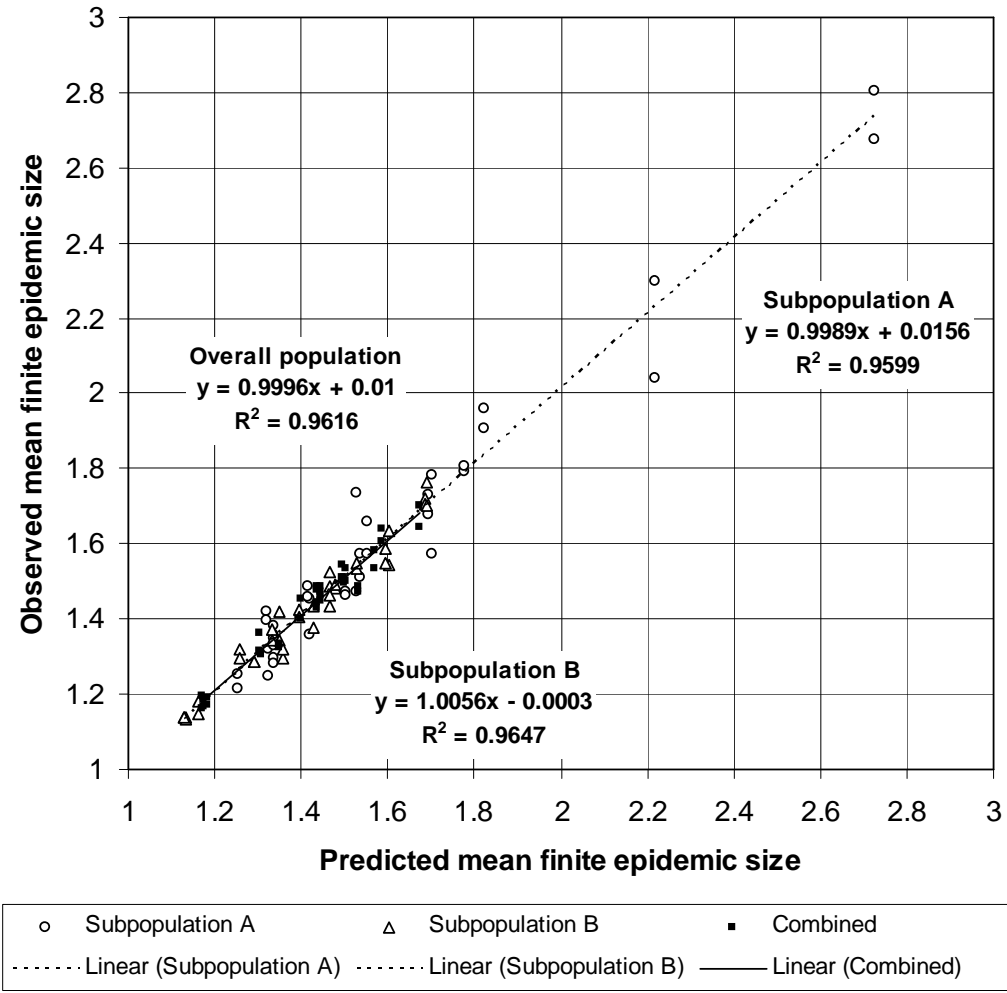}%
\caption{Scatterplot of observed and predicted mean sizes of outbreaks in the
third series of heterogeneous population models. \ Linear regression equations
and $R^{2}$ are shown separately for an initial case randomly chosen from
subpopulation $A$, subpopulation $B$, and from the overall population. \ }%
\label{Finite}%
\end{center}
\end{figure}

Figure \ref{Finite} shows a scatterplot of the observed and predicted mean
size of outbreaks in the third series of heterogeneous population models.
\ The mean size of an outbreak started by a single, randomly chosen imported
infection is $H^{out\prime}(1)$. \ The \textquotedblleft combined" points show
the observed and predicted mean outbreak size for an initial case randomly
chosen from the overall population. \ The mean size of an outbreak started by
an imported infection randomly chosen from subpopulation $A$ is
\[
\frac{\partial}{\partial z}G_{A}(1,H_{f}^{out}(z))|_{z=1,}%
\]
and the mean size of an outbreak started by an imported infection randomly
chosen from subpopulation $B$ is
\[
\frac{\partial}{\partial z}G_{B}(1,H_{f}^{out}(z))|_{z=1}.
\]
The \textquotedblleft subpopulation $A$" points show the observed and
predicted mean outbreak size for an initial case randomly chosen from
subpopulation $A$, and the \textquotedblleft subpopulation $B$" points show
the observed and predicted mean outbreak size for an initial case randomly
chosen from subpopulation $B$. \ All three sets of points are close to the
diagonal, showing that epidemic percolation networks accurately predicted the
mean sizes of finite epidemics in these models. \ Models with higher epidemic
probabilities tend to have smaller outbreak sizes because large outbreaks are
more likely to \textquotedblleft explode" and become epidemics. \ 

\section{Equivalence to classical epidemic theory}

In this section, we show that epidemic percolation networks can reproduce much
of the standard theory of epidemics. \ When the indegree and outdegree are
independent, the epidemic percolation networks predict the same distribution
of outbreak sizes, epidemic threshold, and probability and final size of an
epidemic as the forward and backward branching process approximations.
\ Epidemic percolation networks can also reproduce results from models
developed specifically for special topics within infectious disease
epidemiology. \ Below, we give examples of the derivation of results from the
epidemiology of sexually transmitted diseases, controlled diseases, and
vector-borne diseases. \ 

\subsection{Branching processes}

Much of the mathematical theory of epidemics has been derived using a
branching process as an approximation to the initial spread of disease, where
the \textquotedblleft offspring" of an individual are the persons he or she
infects \cite{Andersson,Diekmann,AndersonMay}. \ The branching process
approximation remains accurate until the first time at which an infectious
individual transmits infection to a person who has already been infected,
which happens after an arbitrarily long interval in the limit of a large
population \cite{Andersson}. \ We will call this the "forward" branching
process approximation to the initial spread of disease, to distinguish it from
the "backward" branching process discussed later.

If the offspring distribution of a branching process has the pgf $g(x)$, then
the probability that the branching process goes extinct is the smallest
solution in $[0,1]$ of the equation $x=g(x)$. \ If $h(z)$ is the pgf for the
total number of individuals generated by the branching process (including the
initial individual), then $h(z)=zg(h(z))$ \cite{Gut}. \ 

\begin{theorem}
\textit{When an epidemic percolation network has independent indegree and
outdegree, it predicts exactly the same outbreak size distribution, epidemic
threshold, and probability of an epidemic (given the infection of a single
randomly chosen node) as a forward branching process approximation to the
initial spread of disease. \ }
\end{theorem}

\begin{proof}
The pgf for the number of secondary infections produced by a randomly chosen
imported infection is $G(1,y)$. \ The number of secondary infections produced
by persons infected through indigenous transmission has the pgf
\[
G_{f}(1,y)=\frac{1}{\left\langle k_{d}\right\rangle }G^{(1,0)}(1,y)=G(1,y),
\]
where the second equality follows from the independence of the indegree and
outdegree. \ Therefore, the initial spread of disease behaves like a branching
process whose offspring distribution has the pgf $g_{f}(y)=G(1,y)$. \ The
probability of a self-limited outbreak (i.e. no epidemic) given the infection
of a randomly chosen individual is the smallest solution in $[0,1]$ of
$y=G(1,y)$, which is equivalent $y=g_{f}(y)$. \ Similarly, $H_{f}%
^{out}(z)=H^{out}(z)$ and the pgf for outbreak sizes is $H^{out}%
(z)=zG(1,H^{out}(z))$, which is equivalent to $h(z)=zg_{f}(h(z))$. \ 
\end{proof}

Another important application of branching processes to SIR models with random
mixing is the use of a "backward" branching process to predict the relative
final size of an epidemic, in which the "offspring" for each individual $i$
are the people who would make infectious contact with $i$ if they were
infectious. \ The relative final size of the epidemic is equal to the
probability that the backward branching process never goes extinct
\cite{Diekmann}. \ 

\begin{theorem}
\textit{When an epidemic percolation network has independent indegree and
outdegree, it predicts exactly the same relative final size of an epidemic as
a backward branching process.}
\end{theorem}

\begin{proof}
In the epidemic percolation network, the number of offspring in the backwards
branching process for a randomly chosen individual has the pgf $G(x,1)$. \ The
pgf for the number of offspring of persons reached by going in reverse along a
directed edge is
\[
G_{r}(x,1)=\frac{1}{\left\langle k_{d}\right\rangle }G^{(0,1)}(x,1)=G(x,1),
\]
where the final equality follows from the independence of the indegree and
outdegree. \ Therefore, the process of moving backwards along edges in the
epidemic percolation network is a branching process whose offspring
distribution has the pgf $g_{b}(x)=G(x,1)$. \ The probability that a node is
not infected in an epidemic is the smallest solution in $[0,1]$ of $x=G(x,1)$,
which equivalent to $x=g_{b}(x)$. \ Therefore, the epidemic percolation
network and the branching process predict the same relative final size of an
epidemic. \ \ 
\end{proof}

In an SIR model whose epidemic percolation network has independent indegree
and outdegree, the epidemic percolation network is a simultaneous mapping of
the forward and backward branching processes. \ However, a branching process
assumes that the offspring distribution is the same in each generation of
infection (because the same pgf $g(x)$ is used for each generation). \ This
assumption fails in an epidemic percolation network in which the indegree and
outdegree are not independent. \ 

Epidemic percolation networks generalize to models with arbitrary joint degree
distributions because they allow the offspring distribution of the initial
node to be different from the offspring distribution of all subsequent
generations in the forward and backward branching processes. \ If we go
forward along edges starting from a randomly chosen node, the offspring
distribution of the initial node has the pgf $G(1,y)$ and the offspring
distribution of nodes in all subsequent generations has the pgf $G_{f}(1,y)$.
\ If we go backward starting from a randomly chosen node, then the offspring
distribution of the initial node has the pgf $G(x,1)$ and all subsequent
generations have the pgf $G_{r}(x,1)$. \ When the indegree and outdegree are
not independent, $G(1,y)\neq G_{f}(1,y)$ and $G(x,1)\neq G_{r}(x,1)$, so both
branching process approximations break down. \ We find it useful to think of
the equations for the component size distributions in Section 3 as describing
(forward or backward) branching processes in which the initial node is allowed
to have a different offspring distribution from all subsequent generations. \ 

By mapping the forward and backward infectious contact processes
simultaneously, the crucial role of the GSCC in the emergence of epidemics
becomes clear. \ In a forthcoming manuscript, we analyze a proportionate
mixing model with three subpopulations: One with the greatest probability of
being in the GIN, one with the greatest probability of being in the GOUT, and
one with the greatest probability of being in the GSCC. \ Vaccinating nodes in
the subpopulation most likely to be in the GSCC is shown to be the most
efficient strategy for reducing both the probability and final size of an
epidemic despite the fact that such nodes were of average infectiousness and
susceptibility. \ We have obtained similar results with network-based models.
\ Nodes with a high probability of being in the GSCC are the \textquotedblleft
core group" that sustains transmission of infection in the population. \ If
the forward and backward infectious contact processes are treated separately,
the notion of the GSCC\ is lost. \ 

\subsection{Other results of epidemic theory}

The use of probability generating functions on epidemic percolation networks
allows many classical results from the theory of epidemics to be re-derived
very easily. \ Below, we give derivations of results from three different
areas of infectious disease epidemiology. \ The ability of epidemic
percolation networks to encompass these results in a single conceptual
framework is a striking demonstration of their utility and generality. \ 

\begin{example}
[Sexually transmitted diseases]For many sexually transmitted diseases,
variation in levels of sexual activity affect the dynamics of disease
transmission. \ One important result is that $R_{0}$ for sexually transmitted
diseases depends on both the mean and the variance of the number of sexual
partners \cite{AndersonMay}. \ Let $I$ be a random variable representing the
expected number of sexual partners a person has during his or her recovery
period. \ If the hazard of infection is proportional to $I$, then%
\[
R_{0}=\frac{T}{\mu_{I}}\left[  \mu_{I}^{2}+\sigma_{I}^{2}\right]  ,
\]
where $\mu_{I}$ and $\sigma_{I}^{2}$ are the mean and variance of the number
of sexual partners and $T$ is the probability of transmission for each
partnership \cite{AndersonMay,Diekmann}. \ 
\end{example}

We first partition the population into subpopulations $1,2,...$ such that
subpopulation $i$ consists of all persons with $I=i$. \ The proportion of the
population in subpopulation $i$ is equal to $P(I=i)$. \ Since there is a
constant probability $T$ of transmission for each partnership and $I=i$ is the
expected number of partners during the recovery period, the mean outdegree of
subpopulation $i$ is $Ti$. \ Since the hazard of infection in subpopulation
$i$ is also proportional to $i$, the mean indegree of subpopulation $i$ must
also be $Ti$. \ The indegree and outdegree of each individual are
conditionally independent given $I$, so the pgf for the degree distribution of
nodes in subpopulation $i$ can be written
\[
G_{i}(x,y)=G_{i}^{in}(x)G_{i}^{out}(y).
\]
The pgf of the degree distribution of the epidemic percolation network is
\[
G(x,y)=\sum_{i}P(I=i)G_{i}^{in}(x)G_{i}^{out}(y),
\]
The overall mean degree is $T\mu_{I}$, and the epidemic threshold is%

\[
\frac{1}{T\mu_{I}}G^{(1,1)}(1,1)=\frac{T}{\mu_{I}}\sum_{i}i^{2}P(I=i)=\frac
{T}{\mu_{I}}\left[  \mu_{I}^{2}+\sigma_{I}^{2}\right]  .
\]
Therefore, the epidemic percolation network correctly predicts the epidemic
threshold for this model. \ However, it can also predict the size distribution
of outbreaks and the probability and final size of an epidemic. \ Epidemic
percolation networks can also be used to analyze SIR\ models with more complex
relationships between sexual activity, infectiousness, and susceptibility.

\begin{example}
[Outbreaks of controlled diseases]For diseases that have been eliminated
within a specific country but are not eradicated worldwide (such as measles in
the United States), imported cases and cases secondary to importation can
still occur. \ To evaluate the success of a disease elimination program, it is
important to determine whether the observed pattern of outbreaks is consistent
with sustained indigenous transmission. \ This can be inferred from the final
size distribution of outbreaks. \ If infectiousness and susceptibility are
independent and cases generate secondary cases according to a Poisson
distribution with mean $R_{0}<1$, all epidemics are finite and epidemic sizes
follow a Borel-Tanner distribution \cite{DeSerres,Haight}. \ The probability
that an outbreak has a final size of $k$ is
\begin{equation}
\frac{R_{0}^{k-1}k^{k-2}e^{-R_{0}k}}{(k-1)!}.\label{S}%
\end{equation}
The pgf for the Borel-Tanner distribution is the unique solution to the equation
\end{example}%

\begin{equation}
h(z)=ze^{R_{0}(h(z)-1)}.\label{Borel}%
\end{equation}

Below the phase transition, the pgf for the distribution of outbreak sizes in
the epidemic percolation network is $H^{out}(z)$. \ Since the incoming and
outgoing degree are independent and the outgoing degree is Poisson distributed
with mean $R_{0}$,%

\[
G_{f}(1,y)=G(1,y)=e^{R_{0}(y-1)}%
\]
But then%

\begin{equation}
H_{f}^{out}(z)=H^{out}(z)=ze^{R_{0}(H^{out}(z)-1)},\label{BTHout}%
\end{equation}
which is identical to equation (\ref{Borel}). \ Therefore, $H^{out}(z)=h(z)$,
so the out-component sizes in the epidemic percolation network have a
Borel-Tanner distribution. \ Using the fact that the probability of having an
outbreak of size one is $e^{-R_{0}}$, it is easy to check that the first few
iterations of equation (\ref{BTHout}) produce coefficients of the form in
equation (\ref{S}).

\begin{example}
[Vector-borne diseases]The model of malaria developed by Ross and Macdonald
consists of humans and mosquitoes. \ Infected humans recover from malaria at a
constant rate $\gamma$, so the average recovery period is $\gamma^{-1}$.
\ There are $m$ susceptible mosquitoes that bite with a rate $a$ and are
infected with probability $c$ when they bite an infectious human, so each
infectious human infects an average of $(amc)\gamma^{-1}$ susceptible
mosquitoes. \ The mortality rate of mosquitoes is $\mu$, so they survive for
an average of $\mu^{-1}$ time units after being infected. \ When an infectious
mosquito bites a susceptible human, the human is infected with probability
$b$, so each infectious mosquito infects an average of $(ab)\mu^{-1}$ humans.
\ The epidemic threshold in this model is defined by%
\[
R_{0}=\frac{ma^{2}bc}{\mu\gamma}.
\]
This was one of the earliest applications of the basic reproductive number
\cite{AndersonMay}.
\end{example}

The full epidemic percolation network for a vector-borne disease would include
nodes representing humans and vectors. \ Humans infect vectors and vectors
infect humans, so every edge in this epidemic percolation network links nodes
of different types. \ Such a network is called a \textit{bipartite network}.
\ Probability generating functions can be used to analyze undirected bipartite
networks \cite{NSW}, and these methods can be adapted to directed bipartite
graphs. \ Using the subscript $h$ for host and $v$ for vector, let
$G_{h}(x,y)$ be the pgf for the degree distribution among humans and let
$G_{v}(x,y)$ be the pgf for the degree distribution among vectors. \ The
epidemic percolation network among humans can then be constructed by drawing
an edge from person $i$ to person $j$ if there is a vector that transfers
infection from $i$ to $j$. \ The pgf for the degree distribution in the
human-human epidemic percolation network is
\[
G_{H}(x,y)=\sum_{j=0}^{\infty}\sum_{k=0}^{\infty}p_{jk}^{h}(G_{vr}%
(x,1))^{j}(G_{vf}(1,y))^{k}=G_{h}(G_{vr}(x,1),G_{vf}(1,y)),
\]
where $p_{jk}^{h}$ is the probability that a human node in the human-vector
epidemic percolation network has $j$ incoming edges and $k$ outgoing edges,
$G_{vr}(x,y)$ is the pgf for the degree of a vector reached by going backwards
along an edge, and $G_{vf}(x,y)$ is the pgf for the degree of a vector reached
by going forward along an edge. \ The outbreak size distribution, the epidemic
threshold, and the probability and final size of an epidemic among humans can
be predicted using $G_{H}(x,y)$. \ \ 

In the Ross-Macdonald model, infectiousness and susceptibility are independent
among both humans and mosquitoes. \ Therefore, $G_{hf}(1,y)=G_{h}(1,y)$ and
$G_{vf}(1,y)=G_{v}(1,y)$. \ The pgf for the out-degree of human nodes is
\[
G_{h}(1,y)=\int_{0}^{\infty}e^{(amc)t(y-1)}\gamma e^{-\gamma t}dt=\frac
{1}{1-amc\gamma^{-1}(y-1)},
\]
and the pgf for the out-degree of mosquito nodes is%
\[
G_{v}(1,y)=\int_{0}^{\infty}e^{(ab)t(y-1)}\mu e^{-\mu t}dt=\frac{1}%
{1-ab\mu^{-1}(y-1)}.
\]
Since the indegree and outdegree are independent in the human-to-human
epidemic percolation network,
\[
G_{Hf}(1,y)=G_{H}(1,y)=G_{h}(1,G_{vf}(1,y)).
\]
Taking the derivative of $G_{Hf}(1,y)$ at $y=1$,
\[
G_{Hf}^{(0,1)}(1,1)=G_{h}^{(0,1)}(1,1)G_{vf}^{(0,1)}(1,1)=\frac{amc}{\gamma
}\cdot\frac{ab}{\mu},
\]
and we see that the epidemic threshold occurs when $\frac{ma^{2}bc}{\mu\gamma
}=1$, which is identical to the threshold derived by Ross and Macdonald. \ \ 

\section{Discussion}

For the epidemic models considered in this paper, methods of finding the exact
distribution of outbreak sizes for a homogeneous population of any fixed size
$n$ exist \cite{Andersson,Diekmann,Gani}. \ However, these methods involve
solving a recursive system of $n$ equations. \ By performing these
calculations in the limit of a large population, the methods presented in this
paper allow a much simpler derivation of the distribution of self-limited
outbreak sizes, the epidemic threshold, and the probability and final size of
an epidemic. \ Our methods also generalize much more easily to heterogeneous
populations. \ 

As proven in the Appendix, the problem of analyzing the final outcomes of any
time-homogeneous stochastic SIR model can be reduced to the problem of
analyzing the components of an epidemic percolation network. \ In
\cite{Kenah}, we showed how epidemic percolation networks can be used to
analyze network-based models of the type studied by Newman \cite{Newman1}.
\ In this paper, we showed that epidemic percolation networks can be used to
analyze stochastic SIR models with random and proportionate mixing. \ In the
limit of a large population, the epidemic percolation network for these models
is purely directed. \ Using the probability generating function for its degree
distribution, we accurately predicted the mean size of outbreaks and the
probability and final size of epidemics for a variety of models in homogeneous
and heterogeneous populations. \ 

The ability of epidemic percolation networks to analyze both network-based and
fully-mixed epidemic models makes them a simple but powerful generalization of
earlier methods of analyzing stochastic SIR models. \ We showed that epidemic
percolation networks with independent indegree and outdegree are equivalent to
forward and backward branching processes, and we used epidemic percolation
networks to re-derive classical results from sexually transmitted diseases,
vector-borne diseases, and controlled diseases. \ Epidemic percolation
networks may also provide a novel and useful qualitative insight into the
control of epidemics. \ The emergence of epidemics corresponds to the
emergence of the GSCC in the epidemic percolation network, so nodes with a
high probability of being in the GSCC may be important targets for
interventions designed to reduce the probability and final size of an
epidemic. \ \ 

\textbf{Acknowledgements:} \textit{This work was supported by the US National
Institutes of Health cooperative agreement 5U01GM076497 "Models of Infectious
Disease Agent Study" (E.K.) and Ruth L. Kirchstein National Research Service
Award 5T32AI007535 "Epidemiology of Infectious Diseases and Biodefense"
(E.K.), as well as a research grant from the Institute for Quantitative Social
Sciences at Harvard University (E.K.). \ We are grateful for the comments of
Marc Lipsitch, James Maguire, Jacco Wallinga, Joel C. Miller, and the
anonymous referees of JTB. \ E.K. would also like to thank Charles Larson and
Stephen P. Luby of the Health Systems and Infectious Diseases Division of
ICDDR,B (Dhaka, Bangladesh).}

\appendix{}

\section{Epidemic percolation networks}

It is possible to define epidemic percolation networks for a much wider range
of stochastic epidemic models than that from the Introduction. \ First, we
specify an SIR epidemic model using probability distributions for infectious
periods in individuals and times from infection to infectious contact in
ordered pairs of individuals. \ Second, we outline time-homogeneity
assumptions under which the epidemic percolation network is defined.
\ Finally, we define infection networks and use them to show that the final
outcome of the epidemic model depends only on the set of initial infections
and the epidemic percolation network. \ This discussion is adapted from that
of our previous paper \cite{Kenah}.

\subsection{Model specification}

Suppose there is a closed population in which every susceptible person is
assigned an index $i\in\{1,...,n\}$. \ A susceptible person is infected upon
infectious contact, and infection leads to recovery with immunity or death.
\ Each person $i$ is infected at his or her \textit{infection time} $t_{i}$,
with $t_{i}=\infty$ if $i$ is never infected. \ Person $i$ is removed (i.e.
recovers from infectiousness or dies) at time $t_{i}+r_{i}$, where the
\textit{recovery period} $r_{i}$ is a random variable with the cumulative
distribution function (cdf) $F_{i}(r)$. \ The recovery period $r_{i}$ may be
the sum of a \textit{latent period}, when $i$ is infected but not yet
infectious, and an \textit{infectious period}, when $i$ can transmit
infection. \ We assume that all infected persons have a finite recovery
period. \ Let $S(t)=\{i:t_{i}>t\}$ be the set of susceptible individuals at
time $t$. \ Let $t_{(1)}\leq t_{(2)}\leq...\leq t_{(n)}$ be the order
statistics of $t_{1},...,t_{n}$, and let $i_{k}$ be the index of the
$k^{\text{th}}$ person infected. \ 

When person $i$ is infected, he or she makes infectious contact with person
$j\neq i$ after an \textit{infectious contact interval} $\tau_{ij}$. \ Given
$r_{i}$, each $\tau_{ij}$ has a conditional cdf $F_{ij}(\tau|r_{i})$. \ Let
$\tau_{ij}=\infty$ if person $i$ never makes infectious contact with person
$j$, so $F_{ij}(\tau|r_{i})$ may have a probability mass concentrated at
infinity. \ Person $i$ cannot transmit disease before being infected or after
recovering from infectiousness, so $F_{ij}(\tau|r_{i})=0$ for all $\tau\leq0$
and $F_{ij}(\tau|r_{i})$ is equal to the conditional probability of
transmission from $i$ to $j$ given $r_{i}$ for all $\tau\in\lbrack
r_{i},\infty)$. \ \ The \textit{infectious contact time} $t_{ij}=t_{i}%
+\tau_{ij}$ is the time at which person $i$ makes infectious contact with
person $j$. \ If person $j$ is susceptible at time $t_{ij}$, then $i$ infects
$j$ and $t_{j}=t_{ij}$. \ If $t_{ij}<\infty$, then we must have $t_{j}\leq
t_{ij}$ because person $j$ avoids infection at $t_{ij}$ only if he or she has
already been infected. \ 

For each person $i$, let his or her \textit{importation time} $t_{0i}$ be the
first time at which he or she experiences infectious contact from outside the
population, with $t_{0i}=\infty$ if this never occurs. \ Let $F_{0}%
(\mathbf{t}_{0})$ be the cdf of the importation time vector $\mathbf{t}%
_{0}=(t_{01},t_{02},...,t_{0n})$.

\subsection{Epidemic algorithm}

Before an epidemic begins, an importation time vector $\mathbf{t}_{0}$ is
chosen. \ The epidemic begins with the introduction of infection at time
$t_{(1)}=\min_{i}(t_{0i})$. \ Person $i_{1}$ is assigned an recovery period
$r_{i_{1}}$. \ Every person $j\in S(t_{(1)})$ is assigned an infectious
contact time $t_{i_{1}j}=t_{(1)}+\tau_{i_{1}j}$. \ We assume that there are no
tied infectious contact times less than infinity. \ The second infection
occurs at $t_{(2)}=\min_{j\in S(t_{(1)})}\min(t_{0j},t_{i_{1}j})$, which is
the time of the first infectious contact after person $i_{1}$ is infected.
\ Person $i_{2}$ is assigned an recovery period $r_{i_{2}}$. \ After the
second infection, each of the remaining susceptibles is assigned an infectious
contact time $t_{i_{2}j}=t_{(2)}+\tau_{i_{2}j}$. \ The third infection occurs
at $t_{(3)}=\min_{j\in S(t_{(2)})}\min(t_{0j},t_{i_{1}j},t_{i_{2}j})$, and so
on.\ \ After $k$ infections, the next infection occurs at $t_{(k+1)}%
=\min_{j\in S(t_{(k)})}\min(t_{0j},t_{i_{1}j},...,t_{i_{k}j})$. \ The epidemic
stops after $m$ infections if and only if $t_{(m+1)}=\infty$. \ 

\subsection{Time homogeneity assumptions}

In principle, the above epidemic algorithm could allow the distributions of
the recovery period and outgoing infectious contact intervals for individual
$i$ to depend on all information about the epidemic available up to time
$t_{i}$. \ In order to generate an epidemic percolation network, we must
ensure that the joint distribution of recovery periods and conditional
transmission probabilities for all ordered pairs of individuals are defined
\textit{a priori}. \ In order to do this, we place the following restrictions
on the model: \ 

\begin{enumerate}
\item We assume that the distribution of the recovery period vector
$\mathbf{r}=(r_{1},r_{2},...,r_{n})$ does not depend on the importation time
vector $\mathbf{t}_{0}$, the contact interval matrix $\mathbf{\tau}=[\tau
_{ij}]$, or the history of the epidemic.

\item We assume that the distribution of the infectious contact interval
matrix $\mathbf{\tau}$ does not depend on $\mathbf{t}_{0}$ or on the history
of the epidemic.
\end{enumerate}

With these time-homogeneity assumptions, the cumulative distribution functions
$F(\mathbf{r})$ of recovery periods and $F(\mathbf{\tau}|\mathbf{r})$ of
infectious contact intervals are defined \textit{a priori}.
\ Given$\ \mathbf{r}$ and $\mathbf{\tau}$, the epidemic percolation network is
a semi-directed network in which there is a directed edge from $i$ to $j$ iff
$\tau_{ij}<\infty$ and $\tau_{ji}=\infty$, a directed edge from $j$ to $i$ iff
$\tau_{ij}=\infty$ and $\tau_{ji}<\infty$, and an undirected edge between $i$
and $j$ iff $\tau_{ij}<\infty$ and $\tau_{ji}<\infty$. \ The entire time
course of the epidemic is determined by $\mathbf{r}$, $\mathbf{\tau}$, and
$\mathbf{t}_{0}$. \ However, its final size depends only on the set
$\{i:t_{0i}<\infty\}$ of imported infections and the epidemic percolation
network. \ In order to prove this, we first define the \textit{infection
network,} which records the chain of infection from a single realization of
the epidemic model.\ 

\subsection{Infection networks}

Let $v_{i}$ be the index of the person who infected person $i$, with $v_{i}=0
$ for imported infections and $v_{i}=\infty$ for uninfected nodes. \ If tied
finite infectious contact times have probability zero, then $v_{i}$ is the
unique $j$ such that $t_{ji}=t_{i}$. \ If tied finite infectious contact times
are possible, then choose $v_{i}$ from all $j$ such that $t_{ji}=t_{i}$. \ The
\textit{infection network} is a network with an edge set $\{v_{i}%
i:0<v_{i}<\infty\}$. \ It is a subgraph of the epidemic percolation network
because $\tau_{v_{i}i}<\infty$ for every edge $v_{i}i$. \ Since each node has
at most one incoming edge, all components of the infection network are trees.
\ Every imported case is either the root node of a tree or an isolated node. \ \ 

The infection network can be represented by a vector $\mathbf{v}%
=(v_{1},..,v_{n})$, which is identical to the "infection network" defined by
Wallinga and Teunis \cite{Wallinga}. \ If $v_{j}=0$, then its infection time
is specified by $\mathbf{t}_{0}$. \ If $j$ was infected through transmission
within the population, then it is connected to an imported infection $imp_{j}
$ in the infection network and its infection time is
\[
t_{j}=t_{imp_{j}}+\sum_{k=1}^{m}\tau_{e_{k}^{\ast}},
\]
where the edges $e_{1}^{\ast},...,e_{m}^{\ast}$ form a directed path from
$imp_{j}$ to $j$. \ This path is unique because all nontrivial components of
the infection network are trees. \ The infection times of all other nodes are
infinite. \ The removal time of each node $i$ is $t_{i}+r_{i}$. \ Therefore,
the entire time course of the epidemic is determined by the importation time
vector $\mathbf{t}_{0}$, the recovery period vector $\mathbf{r}$, the
infectious contact interval matrix $\mathbf{\tau}$. \ \ \ 

\subsection{Final outcomes and percolation networks}

\begin{theorem}
In an epidemic with recovery period vector $\mathbf{r}$ and infectious contact
interval matrix $\mathbf{\tau}$, a node is infected if and only if it is in
the out-component of a node $i$ with $t_{0i}<\infty$ in the epidemic
percolation network. \ Equivalently, a node is infected if and only if its
in-component includes a node $i$ with $t_{0i}<\infty$. \ 
\end{theorem}

\begin{proof}
Suppose that person $j$ is in the out-component of a node $i$ with
$t_{0i}<\infty$ in the epidemic percolation network. \ Then there is a series
of edges $e_{1},...e_{m}$ such that the initial node of $e_{1}$ is $i$, the
terminal node of $e_{m}$ is $j$, and $\tau_{e_{k}}<\infty$ for all $1\leq
k\leq m$. \ Person $j$ receives an infectious contact at or before%
\[
t_{j}^{\ast}=t_{0i}+\sum_{k=1}^{m}\tau_{e_{k}}<\infty,
\]
so $t_{j}\leq t_{j}^{\ast}<\infty$ and $j$ must be infected during the
epidemic. \ To prove the converse, suppose that $t_{j}<\infty$. \ Then there
exists an imported case $i$ and a directed path with edges $e_{1},...,e_{m}$
from $i$ to $j$ such that
\[
t_{j}=t_{i}+\sum_{k=1}^{m}\tau_{e_{k}}.
\]
Since $t_{j}<\infty$, it follows that all $\tau_{e_{k}}<\infty$. \ But then
each $e_{k}$ must be an edge with the proper direction in the epidemic
percolation network, so $j$ is in the out-component of $i$. \ \ \ 
\end{proof}

By the law of iterated expectation (conditioning on $\mathbf{\tau})$, this
result implies that the probability distribution of outbreak sizes caused by
the introduction of infection to node $i$ is identical to that of his or her
out-component sizes in the probability space of epidemic percolation networks.
\ Furthermore, the probability that person $i$ gets infected in an epidemic is
equal to the probability that his or her in-component contains at least one
imported infection. \ In the limit of a large population, the probability that
node $i$ is infected in an epidemic is equal to the probability that he or she
is in the GOUT and the probability that an epidemic results from the infection
of node $i$ is equal to the probability that he or she is in the GIN.

\section{Tables and figures}%

%

\begin{table}[H] \centering
\begin{tabular}
[c]{|l|l|l|l|l|}\hline
\textbf{Distribution } & \textbf{Density function} & \textbf{Support} &
\textbf{Variance} & \textbf{P(}$t\leq.5$\textbf{)}\\\hline
Uniform$(.5,1.5)$ & $1$ & $.5\leq t\leq1.5$ & $.0833$ & $0$\\\hline
$.4+$Gamma$(3,.2)$ & $62.5(t-.4)^{2}e^{-5(t-.4)}$ & $.4\leq t<\infty$ & $.12 $
& $.0143877$\\\hline
$.5+$Exponential$(2)$ & $2e^{-2(t-.5)}$ & $.5\leq t<\infty$ & $.25$ &
$0$\\\hline
Uniform$(0,2)$ & $.5$ & $0\leq t\leq2$ & $.333$ & $.25$\\\hline
Gamma$(2,.5)$ & $4te^{-2t}$ & $0\leq t<\infty$ & $.5$ & $.264241$\\\hline
Exponential$(1)$ & $e^{-t}$ & $0\leq t<\infty$ & $1$ & $.393469$\\\hline
LogNormal$(-.5,1)$ & $\frac{1}{t\sqrt{2\pi}}e^{-.5(.5+\ln t)^{2}}$ & $0\leq
t<\infty$ & $1.71828$ & $.423422$\\\hline
ChiSquare$(1)$ & $\frac{1}{\sqrt{2\pi t}}e^{-.5t}$ & $0\leq t<\infty$ & $2$ &
$.5205$\\\hline
Weibull$(.5,.5)$ & $\frac{1}{\sqrt{2t}}e^{-\sqrt{2t}}$ & $0\leq t<\infty$ &
$5$ & $.632121$\\\hline
Pareto$(.5$,$2)$ & $.5t^{-3}$ & $.5\leq t<\infty$ & $\infty$ & $0$\\\hline
Uniform$(0,4)$* & $.25$ & $0\leq t\leq4$ &  & \\\hline
Exponential$(.5)$* & $.5e^{-.5t}$ & $0\leq t<\infty$ &  & \\\hline
\end{tabular}
\caption{Recovery time distributions for the third series of homogeneous population models.
From top to bottom, they are in order of increasing variance.
The bottom two distributions are used only in the third series of heterogeneous population models.}\label{RecDistributions}%
\end{table}%
%

%

%

%

%

%

%

%

\end{document}